\documentclass[aps,10pt,superscriptaddress,floatfix,nofootinbib,notitlepage]{revtex4-1}

\usepackage{graphicx}
\usepackage{bm} 
\usepackage{amsmath}
\usepackage{amssymb}
\usepackage{braket}
\usepackage{hyperref}
\usepackage{subcaption}

\begin{document}

\title{Quantum Computing for Heavy Quarkonium Spectroscopy}

 \author{Daniel Gallimore}
\email{dpgallim@iu.edu}
\affiliation{ Physics Department and Center for Exploration of Energy and Matter,
Indiana University, 2401 N Milo B. Sampson Lane, Bloomington, IN 47408, USA.} 
\affiliation{Quantum Science and Engineering Center, Indiana University, Bloomington, IN 47408, USA.}

 \author{Jinfeng Liao}
\email{liaoji@indiana.edu}
\affiliation{ Physics Department and Center for Exploration of Energy and Matter,
Indiana University, 2401 N Milo B. Sampson Lane, Bloomington, IN 47408, USA.} 
\affiliation{Quantum Science and Engineering Center, Indiana University, Bloomington, IN 47408, USA.}


\begin{abstract} 

    We report a first demonstration for the application of quantum computing to heavy quarkonium spectroscopy study. Based on a Cornell-potential model for the heavy quark and antiquark system, we show how this Hamiltonian problem can be formulated and solved with the VQE approach on the IBM cloud quantum computing platform. 
     Errors due to a global depolarizing noise channel are corrected with a zero-noise extrapolation method, resulting in good agreement with the expected value. We also generalize the VQE method for solving excited states by orthogonalization  with respect to the ground state. This new approach is demonstrated to be successful for the quarkonium system  on a noiseless quantum simulator and can be easily adapted for solving similar excited state problems in many other physical systems.
\end{abstract}


\maketitle


\section{Introduction}

In recent years, the intersection between quantum computing and nuclear physics has experienced major developments at a rapid pace~\cite{Cloet:2019wre,Zhang_2021,Kharzeev:2021nzh}. While a full simulation of QCD is not yet practical, quantum computers and simulators are currently exploited for  solving/simulating effective models of strong interaction systems as well as related gauge field  theories (e.g. in lower dimensions and/or with smaller symmetry groups)~\cite{Dumitrescu_2018,Lu_2019,Klco:2018kyo,Klco:2018zqz,Roggero:2018hrn,Lee:2019zze,Lamm:2019bik,Alexandru:2019nsa,Ciavarella:2021nmj,Atas_2021,Cohen_2021,Li:2021kcs,Kharzeev:2020kgc,Tu:2019ouv}. 
One category of problem with wide applications to various research fields, such as quantum chemistry and atomic/molecular physics, is computing the energy eigenvalues for a given Hamiltonian. Quantum algorithms for computing eigenvalues mostly come in two flavors, those based on quantum phase estimation (QPE) \cite{kitaev1995quantum,PhysRevLett.83.5162,OBrien_2019} and those based on the variational quantum eigensolver (VQE) \cite{Peruzzo_2014,McClean_2016}. Recently, quantum computations of the ground state energies of few nucleon systems have been achieved using VQE methods \cite{Dumitrescu_2018,Lu_2019}, extending the usefulness of quantum algorithms into the subatomic realm. It is tempting to ask whether quantum computing can be applied to even more fundamental nuclear matter, quarks and antiquarks. Hadron spectroscopy, or how various hadrons are made from their quark/antiquark constituents, is an active research frontier of nuclear physics with many interesting and challenging problems. An example is the heavy quarkonium system for which a Hamiltonian approach with a non-relativistic interaction potential provides a reasonable approximate description. In this work, we perform a first quantum computing study for the ground state as well as excited states of a charm-anticharm system. Our calculation uses the VQE algorithm with unitary coupled cluster (UCC) ansatz \cite{McClean_2016,Shen_2017}. To correct errors due to decoherence in a noisy quantum computer, we further demonstrate a zero-noise extrapolation method for error mitigation. Furthermore, we generalize the VQE method for solving excited states by orthogonalization  with respect to the ground state and demonstrate its success   for the quarkonium system  on a noiseless quantum simulator. The rest of this paper is organized as follows: in Sec. II, the framework of our study will be given, including   the setup of the physics problem and the details of the quantum computation, the variational approach and its generalization to excited states as well as the error mitigation method;   the results of the present study for both ground and excited states will be presented in Sec. III; finally we summarize in Sec. IV. 


\section{Framework}

\subsection{The Physics Problem} 

The physics problem we consider is a pair of charm and anti-charm quarks which form a series of bound states through their mutual interactions. The non-relativistic potential model was shown in past studies to provide a good description of charmonium spectra and the features of such potential were quantitatively determined from phenomenology and lattice calculations~\cite{Bali:2000gf,Kawanai:2011jt}. We will adopt this approach and use the following effective potential
\begin{equation}
    V(r)
    =-\frac{\kappa}{r}+\sigma r,
\end{equation}
known as the Cornell potential. For simplicity, we ignore spin-dependent contributions and consider the above as a spin-averaged potential. We set $\kappa=0.4063$ and $\sqrt{\sigma}=441.6$ MeV, which will result in a ground state energy between that of the physical J/$\psi$ and $\eta_c$. In the center-of-mass frame, the relative motion of the charm and anti-charm is described by the quantum Hamiltonian
\begin{eqnarray} \label{eq_H}
T + V = -\frac{1}{2\mu}\nabla^2 + V (r),
\end{eqnarray}
where $\mu=637.5$ MeV is the reduced mass for the $c$-$\bar{c}$ pair. This defines the problem (i.e. finding eigenvalues and eigenstates of the Hamiltionian) we aim to solve on a quantum computer.

\subsection{Quantum Gate Representation of the Hamiltonian}

Our next step is to represent the Hamiltonian in terms of quantum gate operations that can be implemented on a quantum computer. In comparison to the second-quantized formalism, very little study has been given to preparing first-quantized Hamiltonians on quantum hardware \cite{Tilly_2021}. Thus, we first rewrite eq. (\ref{eq_H}) in second-quantized form:
\begin{equation}\label{eq_H3}
    H_N
    =\sum_{m,n=0}^{N-1}
    \bra{m}
    (T+V)
    \ket{n}
    a_m^\dag
    a_n.
\end{equation} 
While $H_N$ is only exact in the $N\rightarrow\infty$ limit, we must limit $N$ to a finite value since only finitely many orbits can be simulated on a quantum computer at once. By the Hylleraas-Undheim-MacDonald theorem~\cite{HU1930,PhysRev.43.830}, the $n$th eigenvalue of $H_N$ is an upper bound on the $n$th eigenvalue of $H_\infty$. More attention is given to this theorem in the Appendix. The basis $\{\ket{n}\}$ is a set of complete and orthogonal quantum states that spans the Hilbert space for the original physical system under consideration. We take a similar approach to that of \cite{Dumitrescu_2018} by using the spherical quantum harmonic oscillator states as basis orbits. As we are mostly interested in computing the ground state of the Hamiltonian, we will limit ourselves to the $s$-wave states. The operators $a_n^\dag$ and $a_n$ correspond to the creation and annihilation operators for a $c$-$\bar{c}$ pair in the harmonic oscillator $s$-wave state $\ket{n}$. At any point in time, the state of the system in the many-body formalism is of the form $\ket{f_{N-1}\cdots f_1f_0}$, where each $f_n$ represents the number of $c$-$\bar{c}$ pairs in the state $\ket{n}$. Each $\ket{f_n}$ can straightforwardly be identified with a qubit: $\ket{0}=(1,0)^T$ or $\ket{1}=(0,1)^T$. This is possible despite the fact that each $c$-$\bar{c}$ pair is a boson since there is at most one pair per orbit.

The mapping described above is standard for both classical and quantum computational \textit{ab initio} studies of molecular systems that use the popular coupled cluster (CC) and UCC methods. In classical computations, CC is typically preferred since the classical resources needed to implement UCC scale exponentially with system size \cite{Tilly_2021}. However, CC is in general not variational, i.e., convergence to finite energies is not guaranteed \cite{Anand_2022,Tilly_2021}. While possible solutions to this issue do exist, their scaling is also usually exponential \cite{Tilly_2021}. In contrast, UCC is variational, making it immune to explosive failures and a potential attractive alternative to CC on a future fault-tolerant quantum computer. There is also a near-term advantage to using UCC since variational algorithms have been shown to be somewhat resilient to sources of noise present on current quantum devices \cite{Tilly_2021}.

To compute the matrix elements of the Hamiltonian, we use the well-known coordinate-space wavefunctions,
\begin{equation}
    \braket{r|n}
    =(-1)^n
    \sqrt{\frac{2n!}{b^3\,\Gamma(n+3/2)}}
    \exp\!\left(-\frac{r^2}{2b^2}\right)
    L_n^{1/2}
    \!\left(\frac{r^2}{b^2}\right),
\end{equation}
where the oscillator length $b\equiv(\mu\omega)^{-1/2}$ is a function of the oscillator frequency $\omega$ (chosen to be $562.9$ MeV in this calculation) and the reduced mass $\mu$. The matrix elements of the kinetic energy operator are
\begin{equation}
    \bra{m}T\ket{n}
    =\frac{\omega}{2}
    \left\{
    (2n+3/2)\delta_{mn}
    -\sqrt{n(n+1/2)}\,\delta_{m+1,n}
    -\sqrt{(n+1)(n+3/2)}\,\delta_{m-1,n}
    \right\}
\end{equation}
To evaluate the potential energy operator, we separately calculate the matrix elements of $r$ and $r^{-1}$:
\begin{gather}
\bra{m}r\ket{n}
=(-1)^{m+n}
\frac{4b}{\pi(1-4n^2)}
\sqrt{\frac{\Gamma(m+3/2)\Gamma(n+3/2)}{m!n!}}
\,{_2F_1}(2,-m;3/2-n;1),
\\
\bra{m}r^{-1}\ket{n}
=(-1)^{m+n}
\frac{4b^{-1}}{\pi(1+2n)}
\sqrt{\frac{\Gamma(m+3/2)\Gamma(n+3/2)}{m!n!}}
\,{_3F_2}(1/2,1,-m;3/2,1/2-n;1).
\end{gather}
To represent the creation and annihilation operators that previously operated on orbits into quantum gates that act on qubits, we use the Jordan-Wigner transformation \cite{JW_1928},
\begin{align}
    a_n^\dag
    &=\frac{1}{2}
    \left(
    \prod_{j=0}^{n-1}
    Z_j
    \right)
    (X_n-iY_n), \\
    a_n
    &=\frac{1}{2}
    \left(
    \prod_{j=0}^{n-1}
    Z_j
    \right)
    (X_n+iY_n),
\end{align} 
which uses the abbreviated notation $X_n\equiv\sigma_n^x$, $Y_n\equiv\sigma_n^y$, and $Z_n\equiv\sigma_n^z$ for Pauli operators acting on the $n$th qubit. In our simulations, we use a 3-qubit quantum circuit, which computes the Hamiltonian
\begin{gather}
    H_3
    =\sum_{i=0}^9
    H_3^{i}, \\
    H_3^{0}
    =\frac{1}{2}
    \left(
    \frac{21}{4}
    \omega
    +V_{00}
    +V_{11}
    +V_{22}
    \right), \\
    H_3^{1}
    =-\frac{1}{2}
    \left(
    \frac{3}{4}
    \omega
    +V_{00}
    \right)
    Z_0, \label{eq_H31} \\
    H_3^{2}
    =-\frac{1}{2}
    \left(
    \frac{7}{4}
    \omega
    +V_{11}
    \right)
    Z_1, \label{eq_H32} \\
    H_3^{3}
    =-\frac{1}{2}
    \left(
    \frac{11}{4}
    \omega
    +V_{22}
    \right)
    Z_2, \label{eq_H33} \\
    H_3^{4}
    =\frac{1}{4}
    \left(
    -\sqrt{\frac{3}{2}}
    \omega
    +2V_{01}
    \right)
    X_0X_1, \label{eq_H34} \\
    H_3^{5}
    =\frac{1}{4}
    \left(
    -\sqrt{5}
    \omega
    +2V_{12}
    \right)
    X_1X_2, \label{eq_H35} \\
    H_3^{6}
    =\frac{1}{4}
    \left(
    -\sqrt{\frac{3}{2}}
    \omega
    +2V_{01}
    \right)
    Y_0Y_1, \label{eq_H36} \\
    H_3^{7}
    =\frac{1}{4}
    \left(
    -\sqrt{5}
    \omega
    +2V_{12}
    \right)
    Y_1Y_2, \label{eq_H37} \\
    H_3^{8}
    =\frac{1}{2}
    V_{02}
    X_0Z_1X_2, \label{eq_H38} \\
    H_3^{9}
    =\frac{1}{2}
    V_{02}
    Y_0Z_1Y_2, \label{eq_H39}
\end{gather}
with $V_{mn}\equiv\bra{m}V\ket{n}$. Eqs. (\ref{eq_H31})--(\ref{eq_H39}) are proportional to traceless unitary operators.

\subsection{Variational Approach} 

\begin{figure}[!htb]
    \centering
 \includegraphics[width=0.6\textwidth]{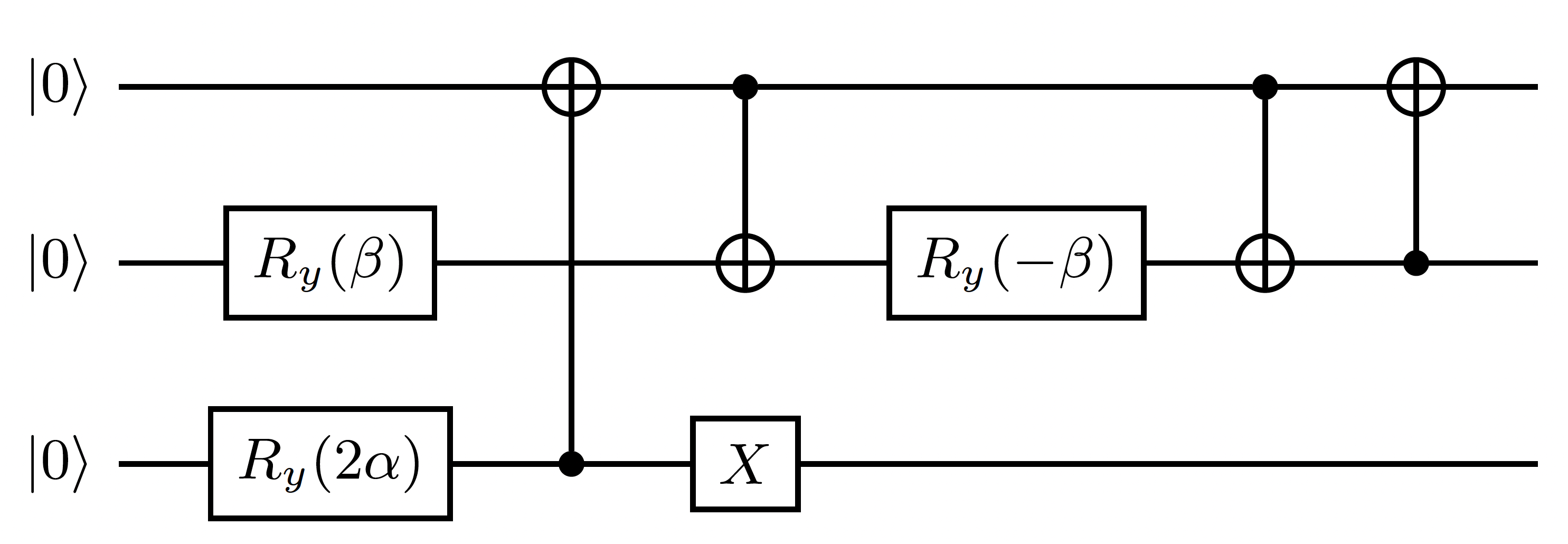}
     \caption{A low-depth gate decomposition of the UCC ansatz used in our simulations. In the diagram, the state is initalized to $\ket{0}$ since this is the initial state encountered on the IBM quantum computers.}
    \label{fig:ansatz}
\end{figure}

The variational principle states that, given an ansatz $\ket{\psi(\vec{\theta})}$ and a Hermitian observable $\mathcal{O}$ that is bounded below, 
\begin{equation}
    \bra{\psi(\vec{\theta})}\mathcal{O}\ket{\psi(\vec{\theta})}
    \geq\epsilon_0,
\end{equation}
where $\epsilon_0$ is the lowest eigenvalue of $\mathcal{O}$. This principle forms the basis of the VQE algorithm, which uses a classical optimization procedure to minimize $\braket{\mathcal{O}}$ with respect to the parameters $\vec{\theta}$ and a quantum subroutine to calculate $\braket{\mathcal{O}}$ for any given $\vec{\theta}$. We approximate the ground state energy of $H_3$ using the VQE algorithm in tandem with the UCC ansatz. For a single $c$-$\bar{c}$ pair with access to three orbitals, this ansatz consists of the unitary operator
\begin{equation}
    U(\theta,\phi)
    =\exp\!\left\{
    \theta
    (a_1^\dag a_0-a_0^\dag a_1)
    +\phi
    (a_2^\dag a_0-a_0^\dag a_2)
    \right\},
\end{equation}
which rotates the state $\ket{001}$ into a linear combination of $\ket{001}$, $\ket{010}$, and $\ket{100}$ with coefficients tuned by $\theta$ and $\phi$. For this specific system, however, it is more convenient to use the parameters $\alpha$ and $\beta$, defined by $\alpha\equiv\sqrt{\theta^2+\phi^2}$ and $\sin\beta\equiv\theta/\alpha$. The 3-qubit UCC ansatz for a single $c$-$\bar{c}$ pair is then just
\begin{equation}\label{eq:3qubitansatz}
    \ket{\psi(\alpha,\beta)}
    =\cos\alpha\ket{001}
    +\sin\alpha\sin\beta\ket{010}
    +\sin\alpha\cos\beta\ket{100}.
\end{equation}
A low-depth gate decomposition of $\ket{\psi(\alpha,\beta)}$ is illustrated in Fig. \ref{fig:ansatz}. 
The variational principle states that for any $\alpha$ and $\beta$,
\begin{equation*}
    \bra{\psi(\alpha,\beta)}
    H_3
    \ket{\psi(\alpha,\beta)}
    \geq
    \epsilon_0,
\end{equation*}
where $\epsilon_0$ is now the ground state energy of $H_3$. To compute $\braket{H_3}$ on a quantum computer with respect to the variational ansatz, we separately measure the traceless unitaries in eqs. (\ref{eq_H31})--(\ref{eq_H39}).

\subsection{Generalized Approximation Scheme for Excited States}

The VQE algorithm has been primarily applied to estimate the ground state energy $\epsilon_0$ and wavefunction 
$\ket{\psi(\alpha_0,\beta_0)}$, which is a kind of limitation for the method. Oftentimes it is interesting and important to also find the excited states of a quantum system, such as the heavy quark spectroscopy problem under consideration in this work. 
Here we generalize the VQE approach to a more comprehensive method for systematically estimating excited states in addition to the ground state. The main idea is to find the next higher energy level (i.e. the 1st excited state) via variational minimization within the sub-Hilbert-space orthogonal to the already found ground state. Obviously this scheme can be carried out further to systematically find the next higher energy level via variational minimization within the sub-Hilbert-space orthogonal to all the lower-lying levels that are already found. The excited states variationally found this way are meaningful as they provide upper bounds on the corresponding true excited state energy values, just like the conventional variational method that gives an upper bound on true ground state energy.   This is based on the so-called Hylleraas-Undheim-MacDonald theorem~\cite{HU1930,PhysRev.43.830}, for which an explicit proof has been included in the Appendix A for readers' convenience.

 To illustrate how this works, let $\ket{\psi(\alpha_1,\beta_1)}$ and $\ket{\psi(\alpha_2,\beta_2)}$ be the first and second excited states with eigenenergies $\epsilon_1$ and $\epsilon_2$ respectively: 
\begin{gather}
   H_3\ket{\psi(\alpha_1,\beta_1)}
    =\epsilon_1  \ket{\psi(\alpha_1,\beta_1)} , \\
  H_3\ket{\psi(\alpha_2,\beta_2)}
    =\epsilon_2   \ket{\psi(\alpha_2,\beta_2)} ,
\end{gather}
where $\epsilon_2>\epsilon_1>\epsilon_0$. Assuming the ground state is known exactly, we consider all possible states that are orthogonal to the ground state, i.e. 
 $\braket{\psi(\alpha_0,\beta_0)|\psi(\alpha,\beta)}=0$. Such states can be expressed in general as linear combinations of all other eigenstates except the ground state. In our case of three basis orbits, we may write 
\begin{equation}
    \ket{\psi(\alpha,\beta)}
    =a\ket{\psi(\alpha_1,\beta_1)}
    +b\ket{\psi(\alpha_2,\beta_2)},
    \quad
    |a|^2+|b^2|=1.
\end{equation}
Therefore,
\begin{equation}
    \bra{\psi(\alpha,\beta)}H_3\ket{\psi(\alpha,\beta)}
    =\epsilon_1
    +|b|^2(\epsilon_2-\epsilon_1)
    \geq
    \epsilon_1
    =\bra{\psi(\alpha_1,\beta_1)}H_3\ket{\psi(\alpha_1,\beta_1)}.
\end{equation}
With an ideal optimization procedure, the first excited state can also be obtained by variational approach in the sub-Hilbert-space orthogonal to the ground state. 

\begin{figure}[!htb]
    \centering
   \includegraphics[width=0.5\textwidth]{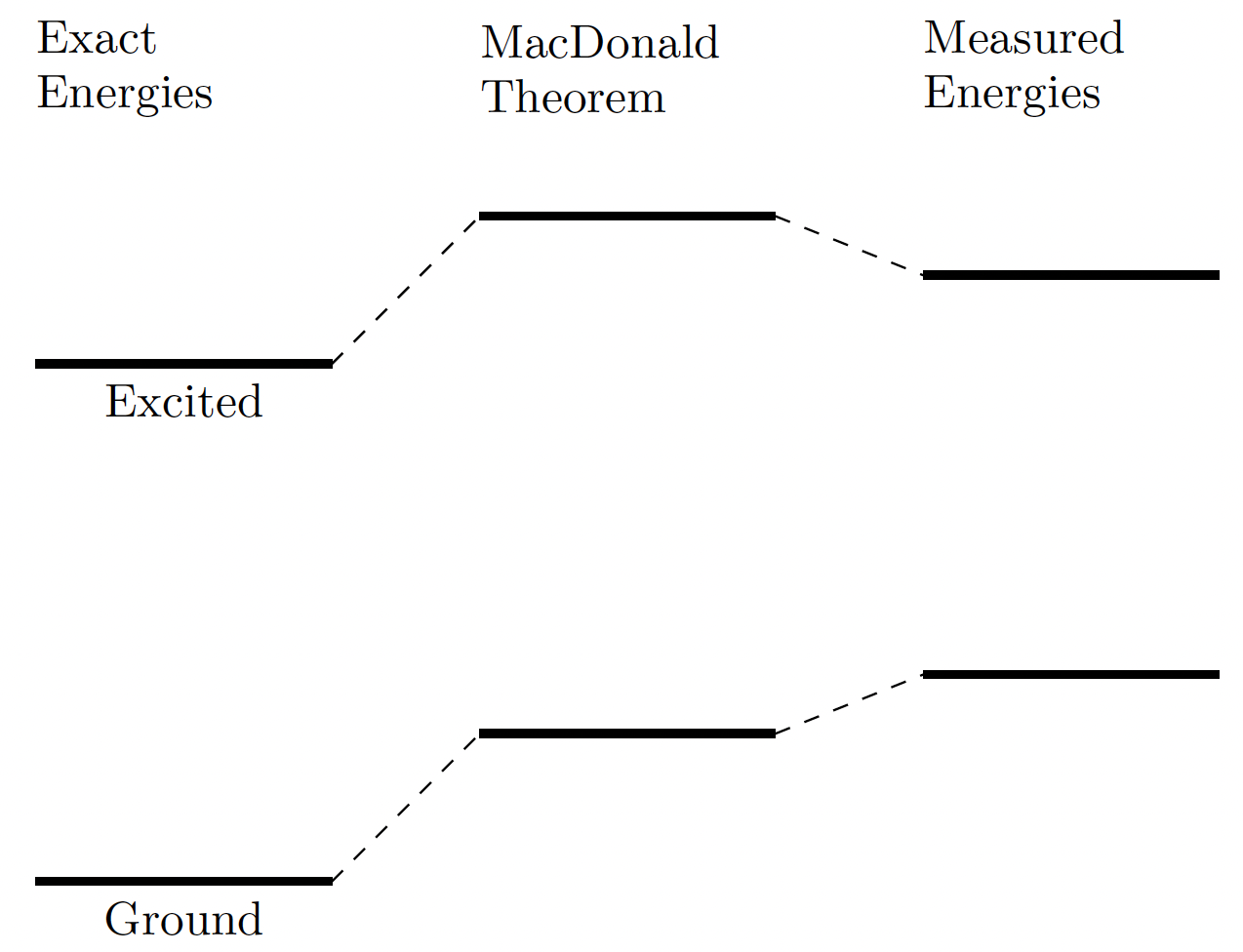}
    \caption{An illustration of the method for estimating excited states. Left column: the lowest two energies of a Hamiltonian with potentially infinitely many energies. Middle column: after truncating to two levels, eigenvalues increase due to the MacDonald theorem. Right column: inaccuracies in variationally determining the ground state cause the ground state measurement to increase and the excited state measurement to decrease. However, the excited state measurement may still be an upper bound on the exact first excited state.}
    \label{fig:levels}
\end{figure}

 Given $\ket{\psi(\alpha_0,\beta_0)}$ is known to good precision, this provides a way to estimate the first excited state by scanning the Hilbert-subspace orthogonal to the ground state and minimizing the expectation value of $H_3$. One can apply this technique iteratively to estimate any excited state energy, with the largest excited state limited by the number of truncated basis orbits. 
 Of course, a tricky issue here is that the ground state itself is obtained via variational method in the first place. So there would be error of the variational ground state with respect to the true ground state. The question is how such error in the ground state would affect the further estimates of excited states. A detailed analysis of the problem, as presented in the Appendix B, concludes that the errors for the excited states stay at the same level as the ground state itself and there will be no worrisome accumulation or even magnification of errors in this method. We also note that even though errors in the calculation of the ground state can cause excited state energy measurements to be less than the eigen-energies of the truncated Hamiltonian, the measurements may still overestimate the eigenenergies of the full Hamiltonian (as illustrated in Fig. \ref{fig:levels}). 
The bottom line is that our method of estimating excited state energies can achieve the same level of accuracy as the conventional VQE method for estimating the ground state. 
\vspace{0.3in} 

\subsection{Error Mitigation}

A key challenge for any quantum calculation is that on a real quantum computer each measurement will necessarily generate errors due to decoherence (environmental noise), such as amplitude damping, phase damping, or depolarizing noise channels. Of these, we choose to correct for a potential global depolarizing channel. Though this channel usually overestimates the degree to which quantum information is lost to the environment, it is appropriate since we have no detailed information about the actual physical noise channel of the quantum computer we are using. To correct for the noise channel, we employ a zero-noise extrapolation method based on \cite{Giurgica_Tiron_2020}. Let
\begin{equation}
    U=L_d\cdots L_2L_1
\end{equation}
be an $N$-qubit quantum circuit with depth $d$. Each layer $L_i$ is composed of one or more quantum gates that can be executed simultaneously. Assuming a global depolarizing channel is the dominant source of noise in an the circuit, the density matrix $\rho$ transforms under $L_i$ in a way that depends only on an ideal noiseless part $\tilde{L}_i$ and a layer-dependent success rate $0\leq r_i\leq1$. That is,
\begin{equation}
    \rho
    \xrightarrow{L_i}
    r_i\tilde{L}_i\rho\tilde{L}_i^\dag
    +\frac{1}{2^N}
    (1-r_i)
    I.
\end{equation}
Consequently, $\rho$ transforms like
\begin{equation}
    \rho
    \xrightarrow{U}
    r\tilde{U}\rho\tilde{U}^\dag
    +\frac{1}{2^N}
    (1-r)
    I
\end{equation}
under the circuit $U$, with total success rate $r\equiv\prod_{i=1}^dr_i$. While $U$ will have a base level of noise that cannot be controlled, it is possible to scale the presence of noise in a predictable manner. Consider the new circuit
\begin{equation}\label{eq:folding}
    V
    \equiv
    U
    (U^\dag U)^n
    (L_1^\dag\cdots L_s^\dag)
    (L_s\cdots L_1),
    \quad
    0\leq s<d.
\end{equation}
While $V$ is logically equivalent to $U$, the ratio of their depths is
\begin{equation}
    \kappa
    \equiv
    2\frac{s}{d}
    +2n+1.
\end{equation}
Under this larger circuit, $\rho$ transforms like
\begin{equation}
    \rho
    \xrightarrow{V}
    r^\lambda\tilde{U}\rho\tilde{U}^\dag
    +\frac{1}{2^N}
    (1-r^\lambda)
    I,
\end{equation}
where
\begin{equation}
    \lambda
    \equiv
    2\frac{\ln q}{\ln r}
    +2n+1
\end{equation}
is a noise scaling parameter and $q\equiv\prod_{i\leq s}r_i$. In the simplest case where $s=0$, the additional noise introduced by $V$ depends only on circuit depth since $\lambda=\kappa$. This is the scaling behavior given the most attention in \cite{Giurgica_Tiron_2020}. However, in the general case where $0\leq s<d$, knowledge of the depths alone is not sufficient. 

Consider a circuit that begins in the pure state $\rho=\ket{\tilde{0}}\!\bra{\tilde{0}}$ and is transformed by the noisy operator $V$. The expectation value of each traceless $H_3^i$ with respect to this state is
\begin{equation}\label{eq:GDscaling}
    \braket{H_3^i}(\lambda)
    =\bra{\tilde{0}}
    \tilde{U}^\dag H_3^i\tilde{U}
    \ket{\tilde{0}}
    r^\lambda,
    \quad
    1\leq i\leq9.
\end{equation}
Evidently, $\braket{H_3^i}(\lambda)$ is proportional to the noiseless expectation value, but vanishes exponentially quickly as $\lambda$ increases beyond $1$. One estimates the noiseless result by measuring $\braket{H_3^i}(\lambda)$ for various $\lambda$, fitting the exponential ansatz to the data, and evaluating the fit at $\lambda=0$. The approach that is simplest and least prone to error is to only gather data for odd $\lambda$. Yet, each time $\lambda$ is increased to the next odd integer, the circuit depth increases by $2d$. After only a few values of $\lambda$, the depth may be too large for a given quantum processor to handle without introducing significant errors. To build a circuit with arbitrary $\lambda\geq1$, resulting in a better fit, one needs precise knowledge of each $r_i$, which is impractical for even moderately large circuits. However, under the simplifying assumption that each $r_i$ is approximately equal, $\lambda\approx\kappa$. In other words, one can approximate the true scaling behavior using only circuit depths.

\begin{figure}[!htb]
\begin{subfigure}[t]{0.3\textwidth}
    \includegraphics[width=\linewidth]{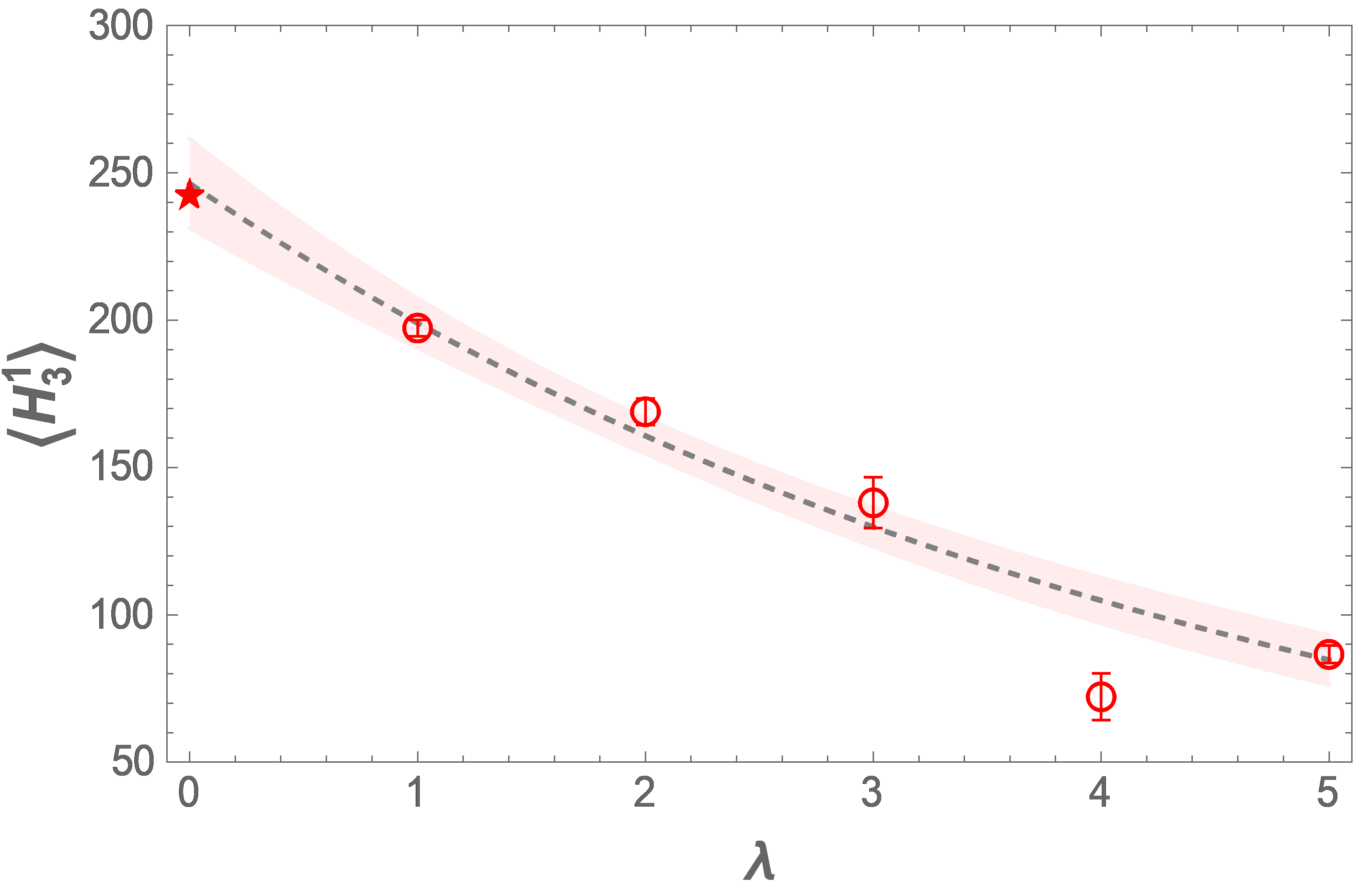}
\end{subfigure}\hfill
\begin{subfigure}[t]{0.3\textwidth}
  \includegraphics[width=\linewidth]{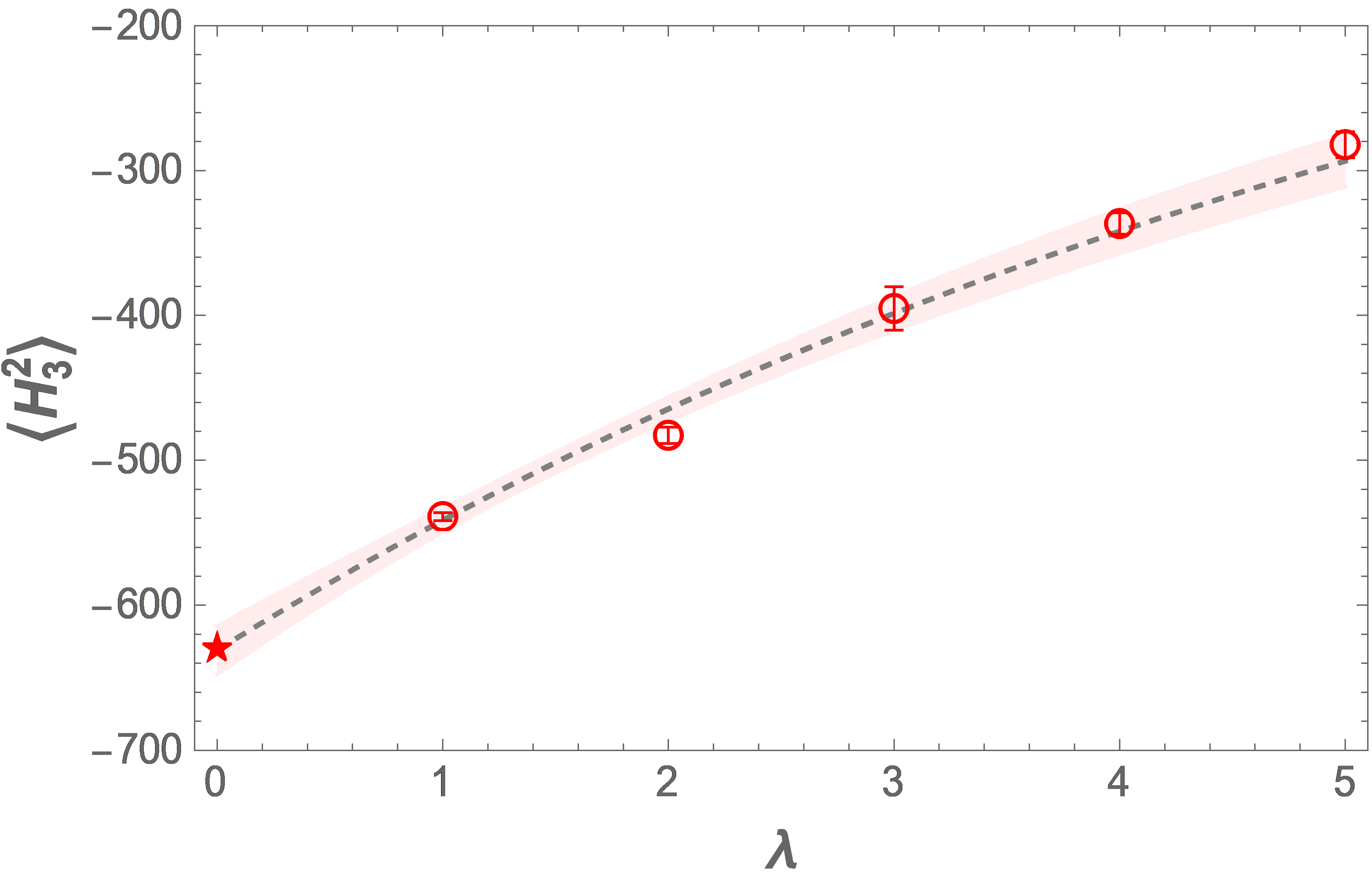}
\end{subfigure}\hfill
\begin{subfigure}[t]{0.3\textwidth}
    \includegraphics[width=\linewidth]{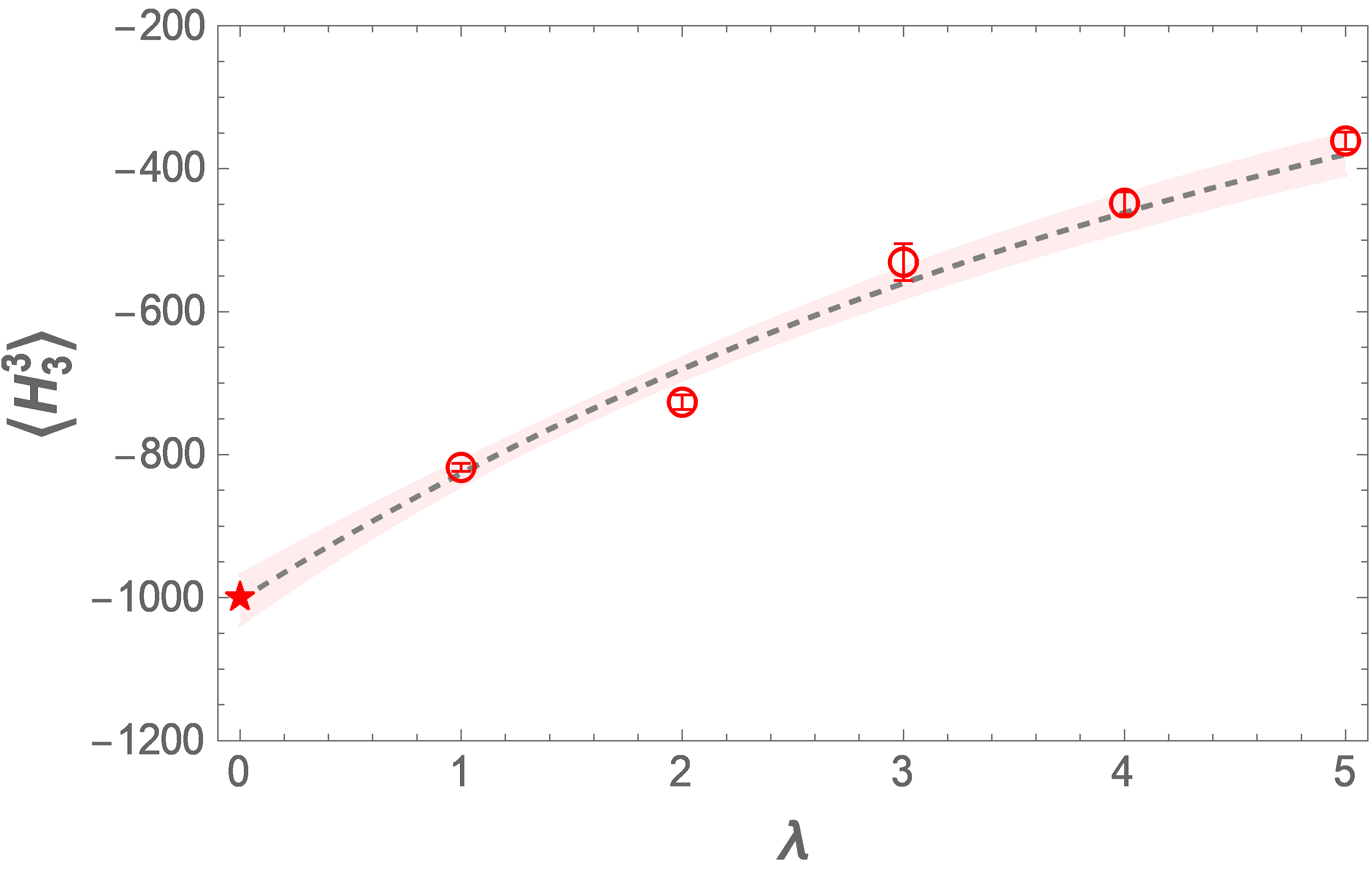}
\end{subfigure}

\begin{subfigure}[t]{0.3\textwidth}
    \includegraphics[width=\linewidth]{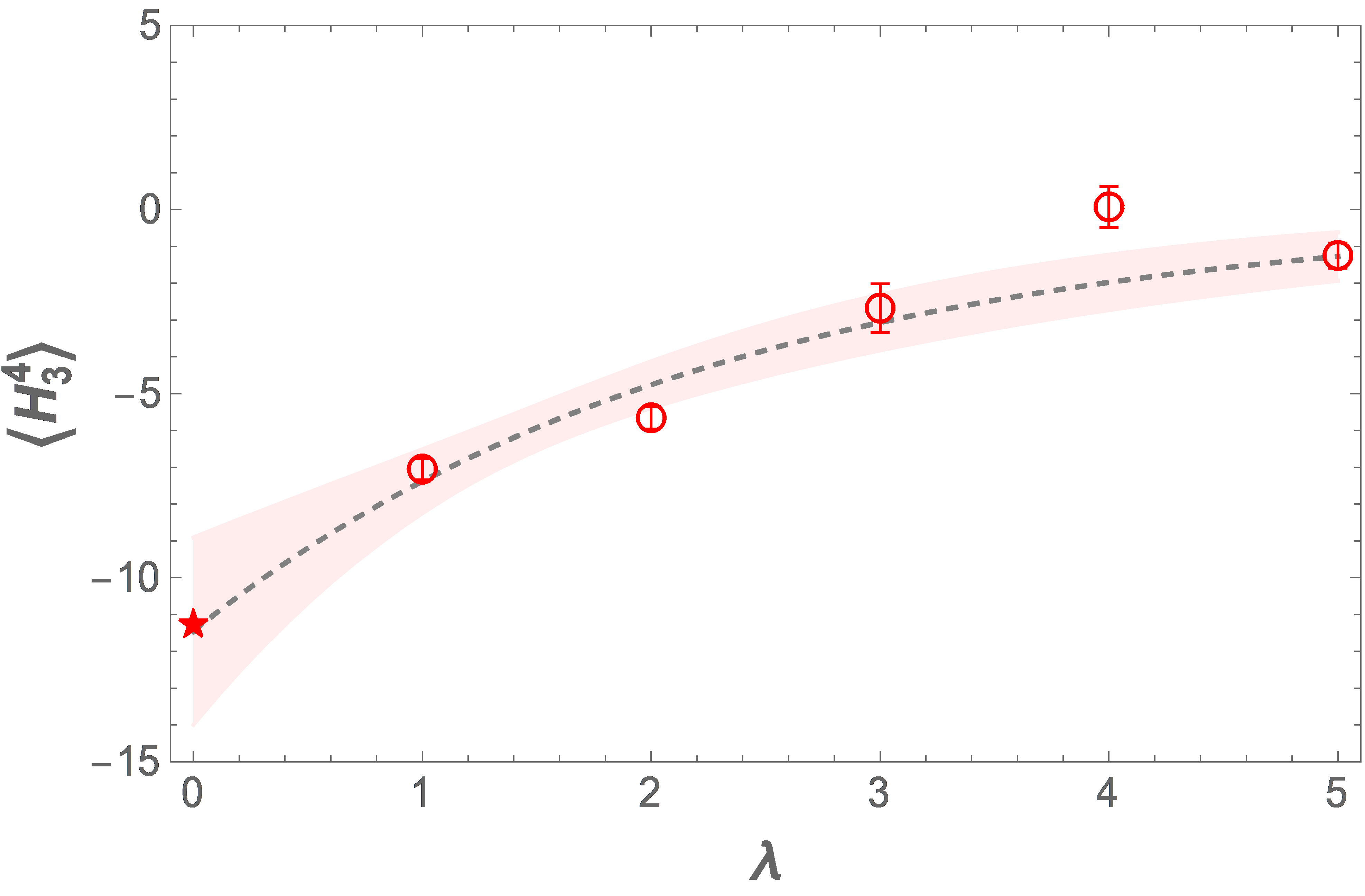}
\end{subfigure}\hfill
\begin{subfigure}[t]{0.3\textwidth}
    \includegraphics[width=\linewidth]{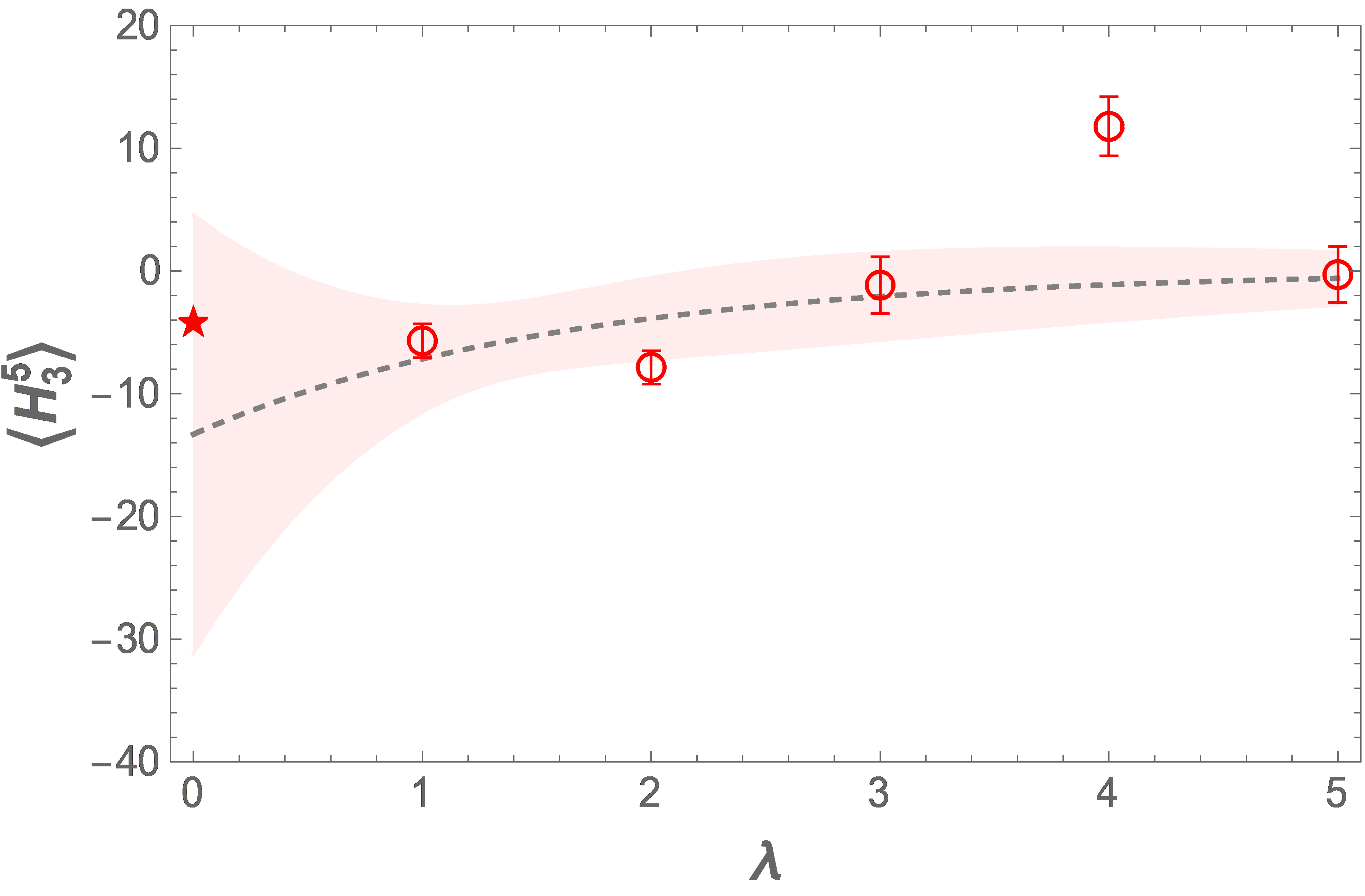}
\end{subfigure}\hfill
\begin{subfigure}[t]{0.3\textwidth}
    \includegraphics[width=\textwidth]{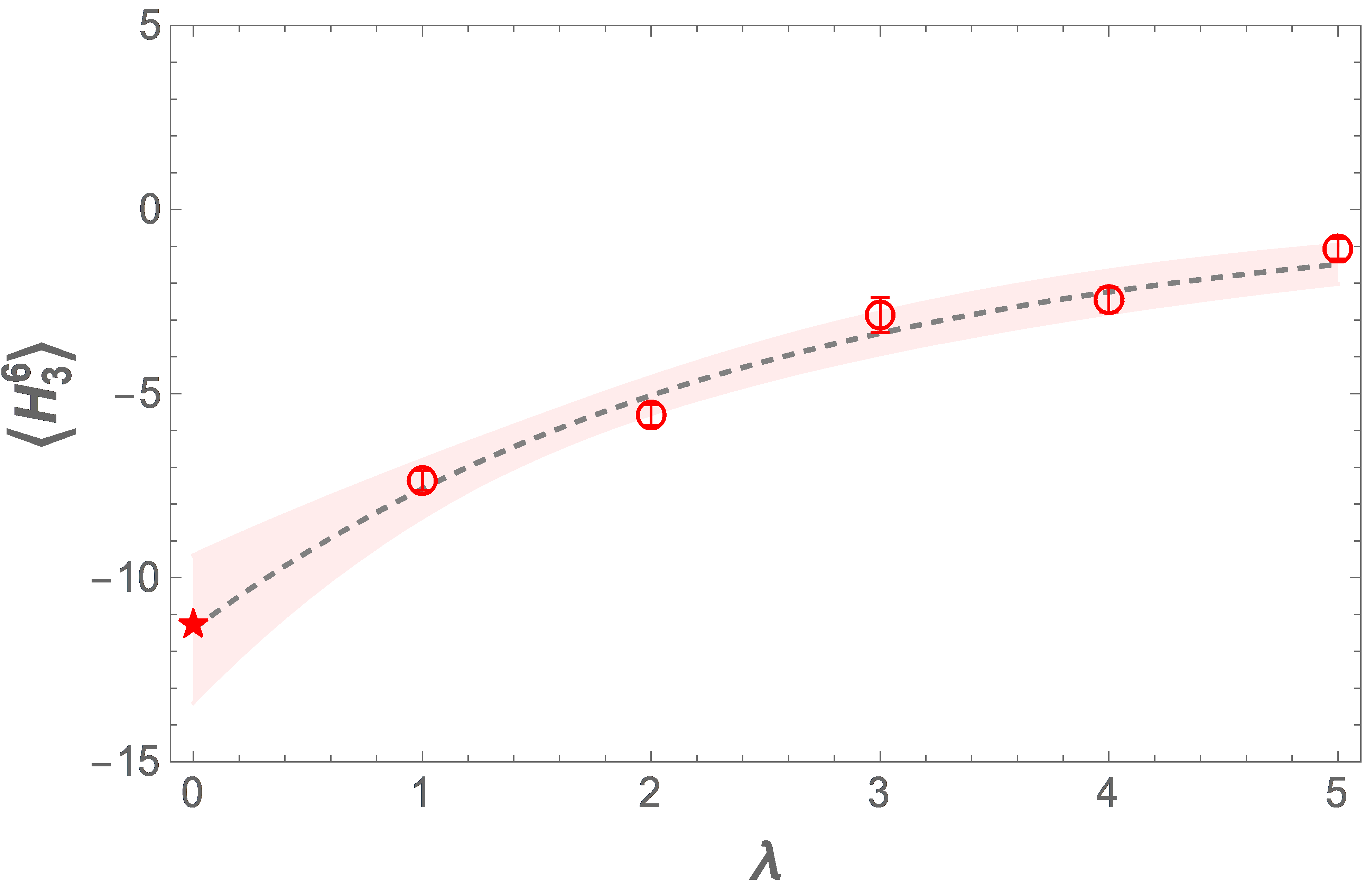}
\end{subfigure}

\begin{subfigure}[t]{0.3\textwidth}
    \includegraphics[width=\linewidth]{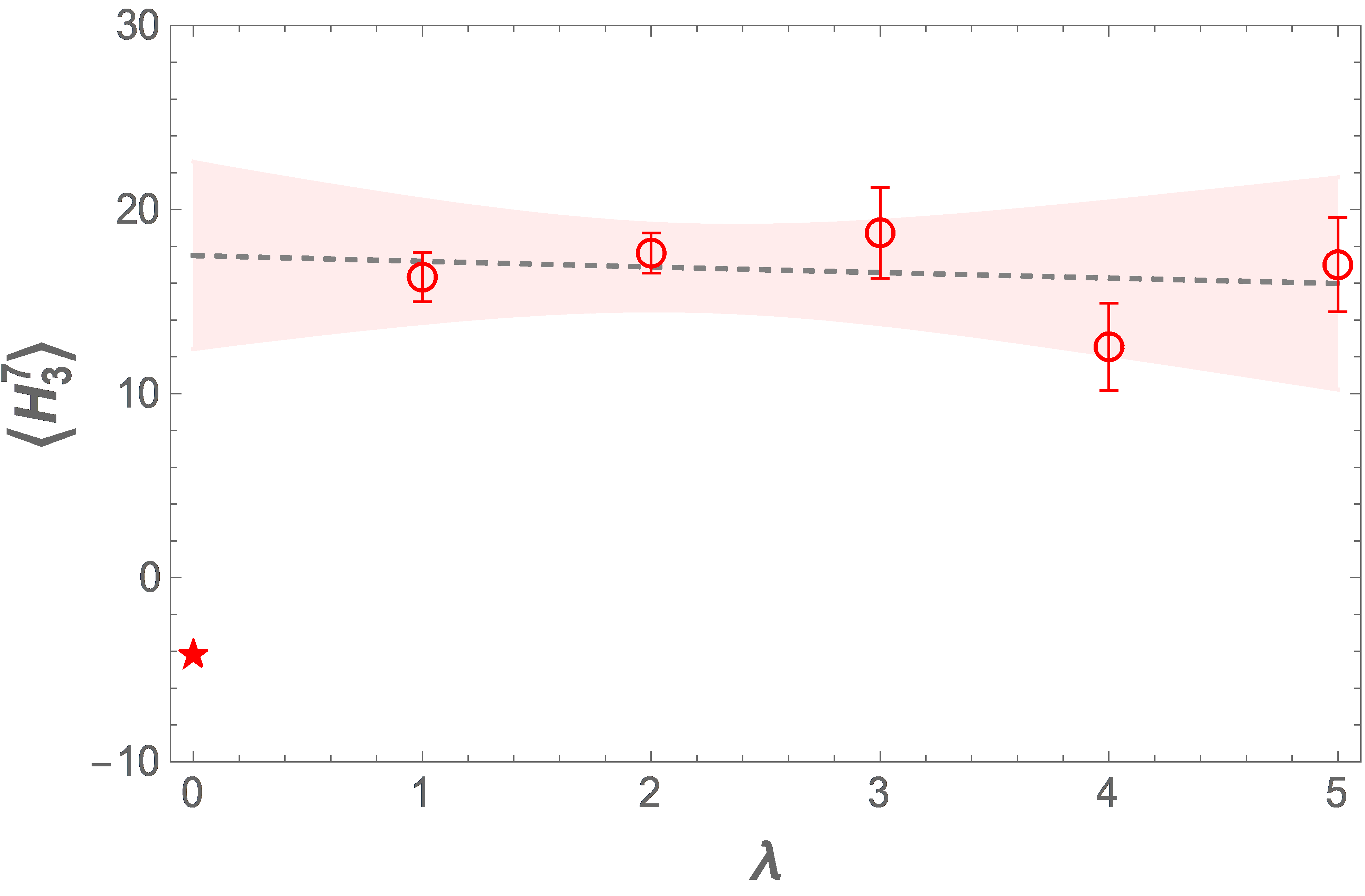}
\end{subfigure}\hfill
\begin{subfigure}[t]{0.3\textwidth}
    \includegraphics[width=\linewidth]{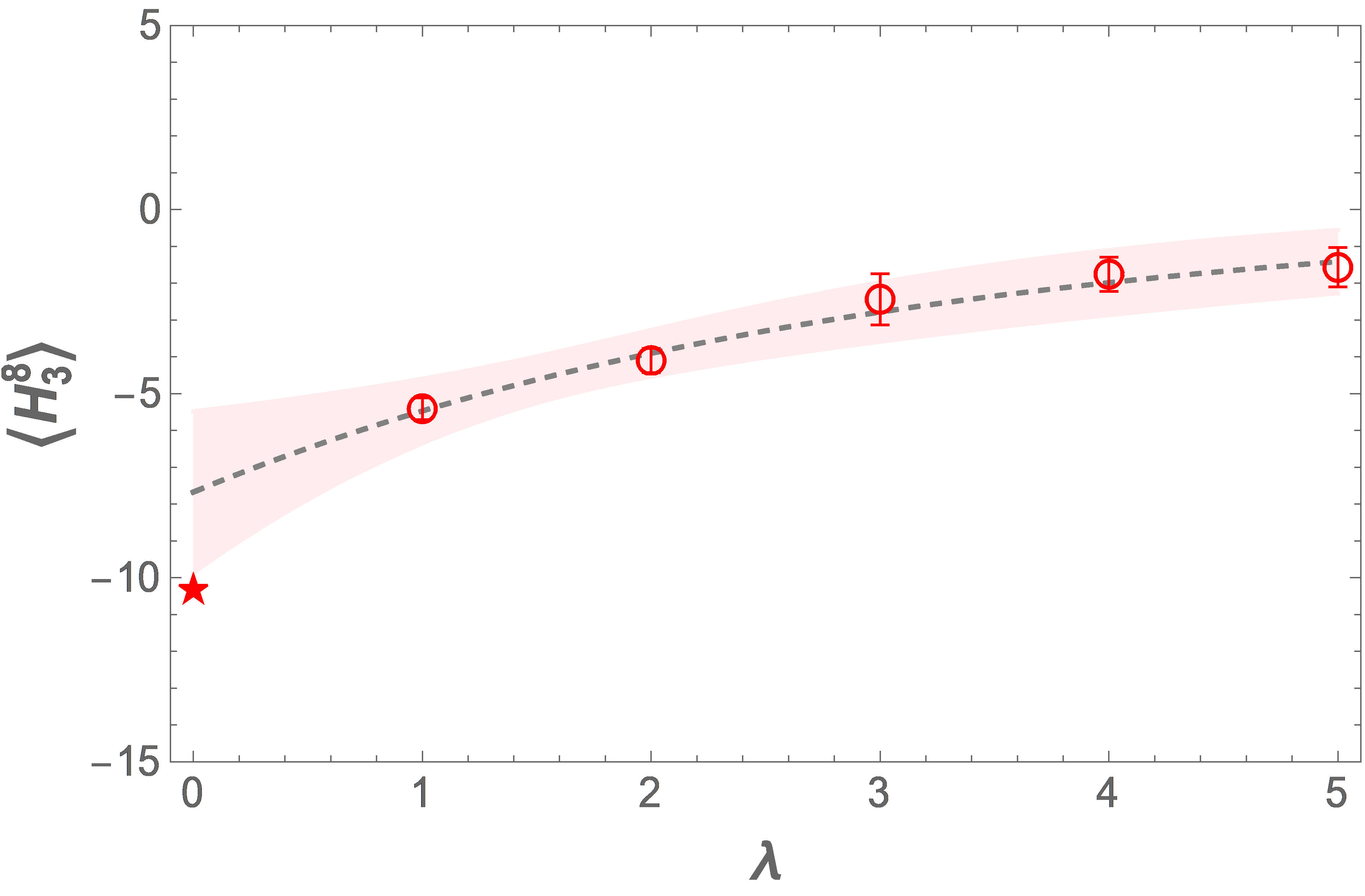}
\end{subfigure}\hfill
\begin{subfigure}[t]{0.3\textwidth}
    \includegraphics[width=\linewidth]{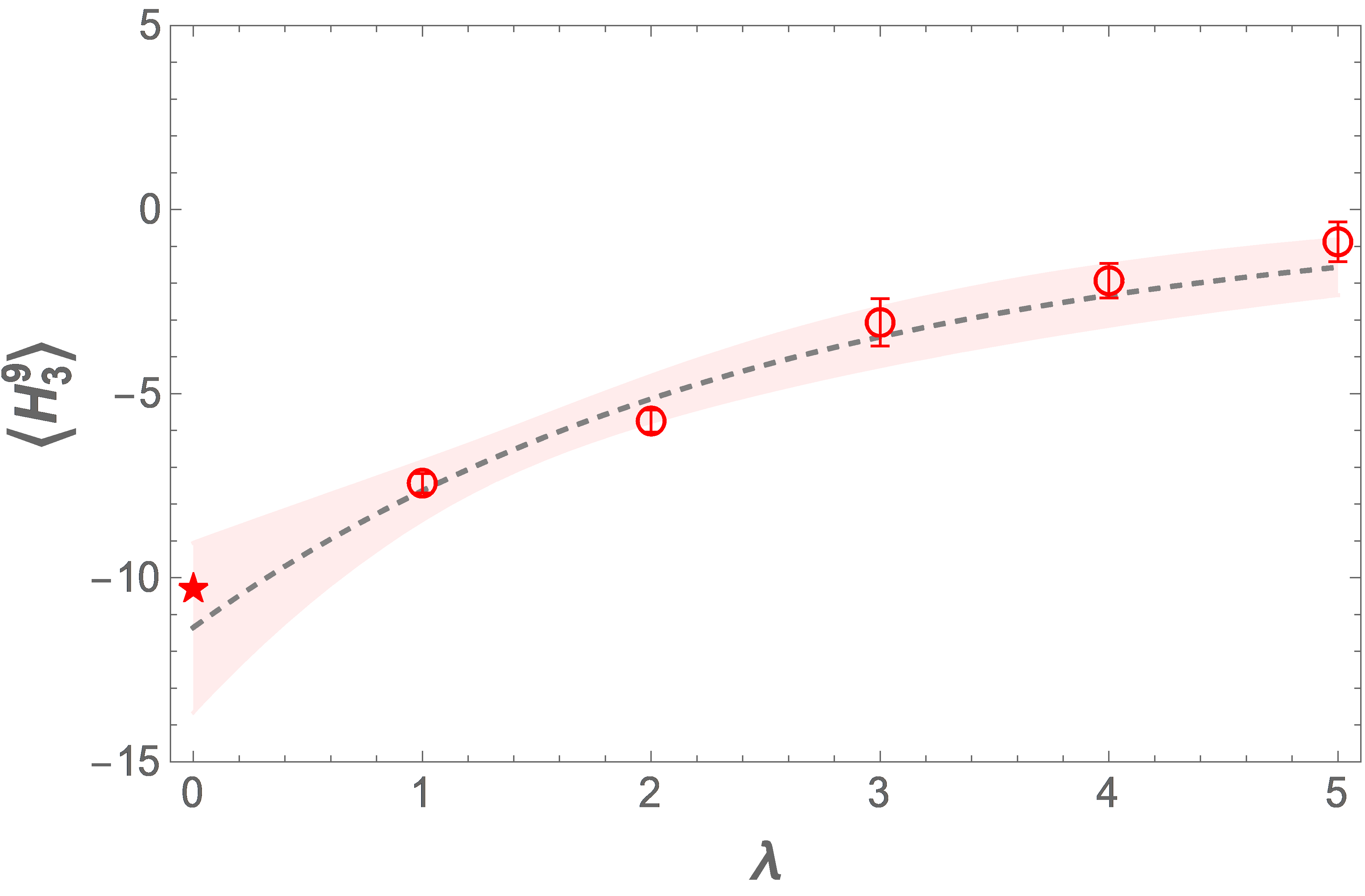}
\end{subfigure}

\caption{Expectation values (in Mev) of eqs. (\ref{eq_H31})--(\ref{eq_H39}) versus the scaling parameter $\lambda$ with $2\sigma$ prediction bands. Actual values have $95\%$ chance of lying within bands. Noiseless quantum simulation results are indicated by stars, but are not included in the fits.}
\label{fig:unitaries}
\end{figure}

\subsection{Scaling}

While the current study is based on three-qubit circuits representing three physical states, it is useful to think about the scaling aspect when performing a similar calculation with $N$ physical states, with each quantum state represented by one qubit as in typical VQE approach. The number of Pauli strings created by the Jordan-Wigner transformation is $O(N^4)$ and computing the expectation value of each Pauli string at precision $\epsilon$ requires $O(1/\epsilon^2)$ repetitions of the circuit. Thus, $O(N^4/\epsilon^2)$ total circuits must be evaluated. The depth of each evaluation depends on the quantum gate representation of the UCC ansatz. While it is possible to find low-depth representations of the ansatz for small systems through experimentation, the standard approach for larger systems is to rewrite the ansatz using a Suzuki-Trotter decomposition. For $k$ Trotter steps, each with a depth $O(N^2)$, the total depth of a single evaluation is $O(kN^2)$. However, for an accurate description of the ground state in simple systems, it is often sufficient to set $k=1$ \cite{Tilly_2021}. Additionally, when using a real quantum computer, noise mitigation will further increase the circuit depth. The zero-noise extrapolation method described above increases the depth of a given evaluation by a factor $\lambda$.

\section{Results}

\subsection{The Ground State} 

\begin{figure}[!htb]
    \centering
    \includegraphics[width=0.6\textwidth]{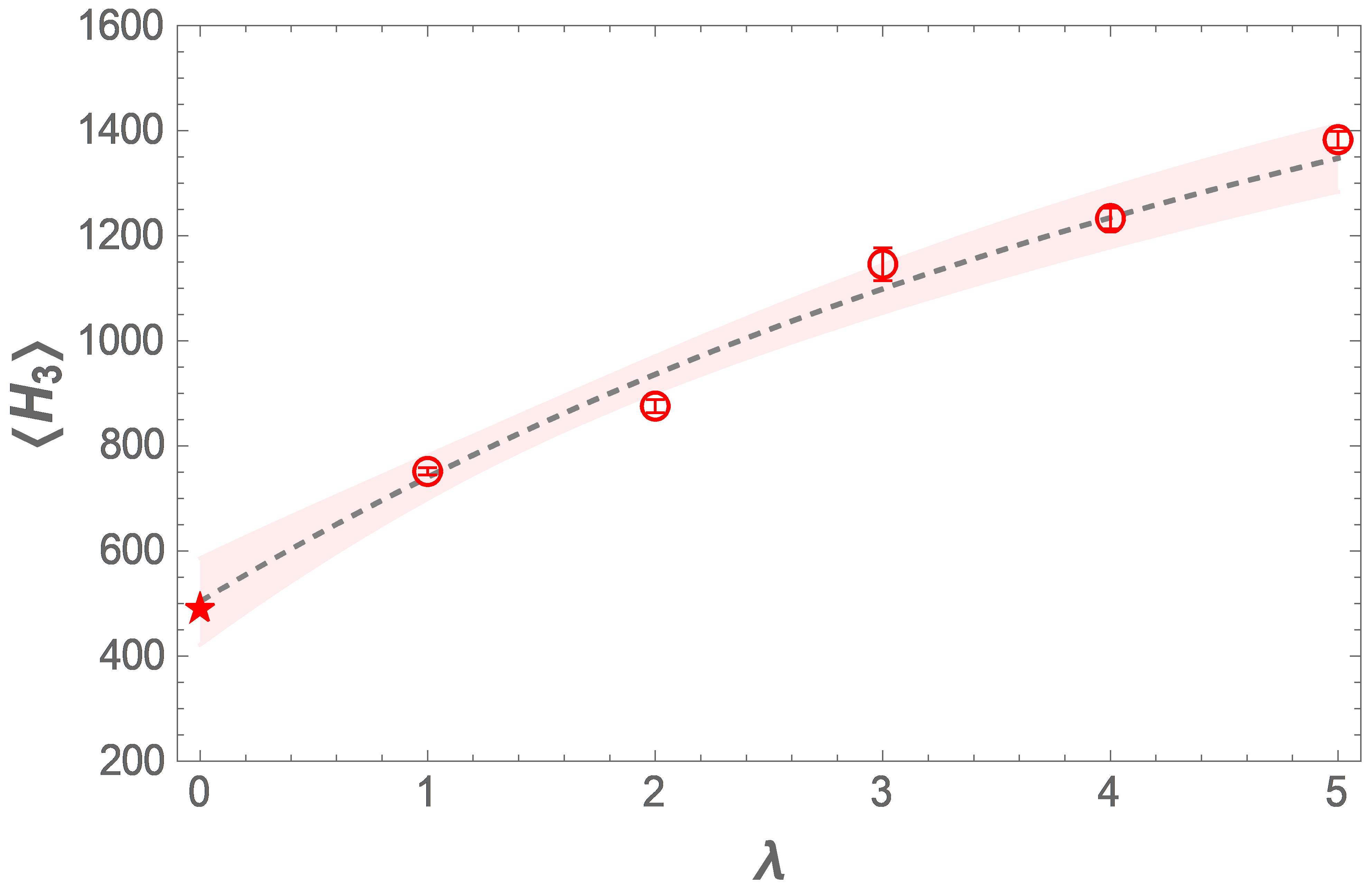}
    \caption{Expectation value (in MeV) of $H_3$, obtained by combining the plots in Fig. \ref{fig:unitaries} with $2\sigma$ prediction bands. Actual values have $95\%$ chance of lying within bands. Noiseless quantum simulation result is indicated by the star symbol for comparison.}
    \label{fig:H3}
\end{figure}

For this work, we used the IBM's cloud quantum computing platform and   ran our circuit on IBMQ Athens, which uses a 5-qubit Falcon r4 quantum processor. The IBM platform also provides the QASM Simulator that one can use to generate quantum calculation results in an ideal noiseless setting. To calculate the expectation value of $H_3$ for a given $\alpha,\beta,\lambda$, eqs. (\ref{eq_H31})--(\ref{eq_H39}) were measured separately $1.024\times10^6$ times. Beginning with $\lambda=1$, we used the VQE algorithm to find the appropriate $\alpha_0$ and $\beta_0$ corresponding to the ground state $\ket{\psi_\text{1S}}$. These were determined to be $\alpha_0=3.31$ and $\beta_0=0.95$. We then calculated the expectation values of eqs. (\ref{eq_H31})--(\ref{eq_H39}) with respect to $\ket{\psi_\text{1S}}$ for $\lambda=2,3,4,5$. These data are shown in Fig. \ref{fig:unitaries}. As one can see, overall we've found quite reasonable scaling behavior in line with expectations and the extrapolation results toward the $\lambda=0$ limit are in good agreement with noiseless results for most cases. However, the plots for $\braket{H_3^5}$ and $\braket{H_3^7}$ conform especially poorly to the global depolarizing model. In these cases, a different noise model may be needed to reduce the uncertainty in the extrapolated value. Combining these data gives the final plot for $\braket{H_3}(\lambda)$, shown in Fig. \ref{fig:H3}. Here we also list the obtained  values:  
$\braket{H_3}(\lambda=1) = 751.57 \pm 7.20 \ \rm MeV $; 
$\braket{H_3}(\lambda=2) = 875.87 \pm 12.41 \ \rm MeV $; 
$\braket{H_3}(\lambda=3) = 1146.05 \pm 31.13 \ \rm MeV $; 
$\braket{H_3}(\lambda=4) = 1232.98 \pm 19.63 \ \rm MeV $; 
$\braket{H_3}(\lambda=5) = 1382.92 \pm 15.83 \ \rm MeV $.
As one can see, the statistical errors of each calculation at given $\lambda$ are rather small, mostly at $(1\sim 2)\%$ level.

 The extrapolation toward noiseless limit gives a value of $\braket{H_3}(\lambda \to 0) = 502\pm98 \ \rm MeV$. The corresponding ground state wavefunction is found to be
\begin{equation}
    \ket{\psi_\text{1S}}
    =-0.9858\ket{001}
    -0.1369\ket{010}
    -0.09722\ket{100}.
\end{equation}
Notice that $\ket{\psi_\text{1S}}$ is almost entirely composed of the harmonic oscillator ground state, supporting our choice of basis.

The central value of $\braket{H_3}$ from such extrapolation, $502\ \rm MeV$, compares well with both the result of $493\pm1 \ \rm MeV$ from the  noiseless QASM Simulator and the expected value of $492.6 \ \rm MeV$ from exact  diagonalization of $H_3$. 
 The $\pm 98 \ \rm MeV$ error   represents a $2\sigma$ uncertainty band dominantly from the extrapolation uncertainty.  While   the quantum algorithm itself generates rather small errors, there is still   sizable uncertainty due to the extrapolation, which is actually a useful reflection of the limitation due to noisy quantum computers. On such real-world devices, the actual noisy behaviors could go beyond the strategy we adopt for error mitigation  while only a perfect understanding of noise sources could help substantially reduce the extrapolation uncertainty.

\subsection{Estimating an Excited State}

Next, we calculate the first excited state energy of the charmonium system under consideration. There are several effective methods for estimating excited state energies using a variational algorithm \cite{Wang_1994,McClean_2016,Shen_2017,Santagati_2018,Lee_2019}, some of which were developed before their applications to quantum computing were realized. In this work, we measure the 2S noiseless expectation value of $H_3$ by orthogonalizing the UCC ansatz with respect to our estimate of the ground state, then applying the VQE algorithm to this reduced Hilbert space. This approach is likely to be effective since we only use the noiseless QASM Simulator, which will keep statistical errors small. Due to the orthogonalization constraint, the $\alpha_1$ and $\beta_1$ are not independent. Basic geometric considerations suggest that $\alpha_1$ and $\beta_1$ be related by a third parameter $\gamma$:
\begin{gather}
    \cos\alpha_1
    =-\sin\alpha_0\cos\gamma, \\
    \sin\alpha_1\sin\beta_1
    =\cos\alpha_0\sin\beta_0\cos\gamma
    +\cos\beta_0\sin\gamma, \\
    \sin\alpha_1\cos\beta_1
    =\cos\alpha_0\cos\beta_0\cos\gamma
    -\sin\beta_0\sin\gamma.
\end{gather}
We find the 2S energy to be $1212\pm2$ MeV, which compares well with the expected value of $1210.8 \ \rm MeV$ from exact diagonalization. The corresponding wavefunction is
\begin{equation}
    \ket{\psi_\text{2S}}
    =-0.1617\ket{001}
    +0.9298\ket{010}
    +0.3307\ket{100},
\end{equation}
with $\gamma=2.87$. By employing more qubits for the computation, one can easily extend this strategy to calculate higher and higher excited states.

\section{Summary}

In summary,  we have reported a first demonstration for the application of quantum computing to heavy quarkonium spectroscopy study. Based on a Cornell-potential model for the heavy quark and antiquark system, we've shown how this Hamiltonian problem can be formulated and solved with the VQE approach on the IBM cloud quantum computing platform. Errors due to a global depolarizing noise channel on a real quantum computer have been mitigated with a zero-noise extrapolation method, resulting in good agreement with the expected value for the ground state. We've also generalized the VQE method for solving excited states by orthogonalization  with respect to the ground state and analyzed the error in such estimates of excited states. This new method has been successfully demonstrated for the quarkonium system  on a noiseless quantum simulator and shall be generally applicable for solving similar excited state problems in many other physical systems. With the current explorative study  showing the potential of quantum computing for quark dynamics in hadron spectroscopy, it is tempting to fully exploit the possibility of solving more challenging problems in this area (such as exotic states) on a quantum computer, which will be our future work to be reported elsewhere.

\vspace{0.2in}
\section*{Acknowledgments} 
We are very grateful to G. Ortiz for many valuable discussions, in particular on the variational methods for excited states.  
We also thank M. Shepherd and A. Szczepaniak  for  helpful comments. This work is supported by  the NSF Grant No. PHY-2209183. This material is partly based upon work done within the framework of the ExoHad Collaboration, supported by the U.S. Department of Energy, Office of Science, Office of Nuclear Physics.

\section{Appendix}

\subsection{The Hylleraas-Undheim-MacDonald Theorem}

The following is a proof of the Hylleraas-Undheim-MacDonald theorem~\cite{HU1930,PhysRev.43.830} for first-quantized Hamiltonians in a simple setting applicable to this research. The result is immediately applicable to second-quantized Hamiltonians since the eigenvalues are unchanged. Suppose a given Hamiltonian $H$ operating on the Hilbert space $\mathcal{H}$ has the matrix representation
\begin{equation}
    H
    =\begin{pmatrix}
        h & X \\
        X^\dag & Y
    \end{pmatrix},
\end{equation}
where $h=h^\dag$ and $Y=Y^\dag$. The submatrices $X$ and $Y$ may be infinite; however, the $N\times N$ block $h$ is finite. Let the normalized eigenvectors of $h$ satisfy
\begin{equation}
    h\ket{\varphi_n}
    =\lambda_n\ket{\varphi_n},
    \quad
    \lambda_0<\lambda_1<\cdots<\lambda_{N-1}.
\end{equation}
Define the states $\ket{\tilde{\varphi}_n}\equiv(\ket{\varphi_n},0)\in\mathcal{H}$, with $n=0,\cdots,N-1$. Then
\begin{equation}
    \bra{\tilde{\varphi}_n}H\ket{\tilde{\varphi}_n}
    =\bra{\varphi_n}h\ket{\varphi_n}
    =\lambda_n.
\end{equation}
Next, let the normalized eigenvectors of $H$ satisfy
\begin{equation}
    H\ket{v_i}
    =\Lambda_i\ket{v_i},
    \quad
    \Lambda_0<\Lambda_1<\cdots,
\end{equation}
where the spectrum of $H$ may be infinite. Consider the ground state $\ket{v_0}$. Since $\ket{v_0}$ minimizes the Rayleigh quotient of $H$,
\begin{equation}
    \Lambda_0
    =\bra{v_0}H\ket{v_0}
    \leq
    \bra{\tilde{\varphi}_0}H\ket{\tilde{\varphi}_0}
    =\lambda_0,
\end{equation}
that is, $\Lambda_0\leq\lambda_0$. Now, define an auxiliary  state that is orthogonal to the ground state: 
\begin{equation}
    \ket{\Phi_1}
    =c_{01}
    \ket{\tilde{\varphi}_0}
    +c_{11}
    \ket{\tilde{\varphi}_1}
\end{equation}
such that $\braket{v_0|\Phi_1}=0$ and $\braket{\Phi_1|\Phi_1}=1$. Then
\begin{equation} \label{eq_51}
  \Lambda_1 =  \bra{v_1}H\ket{v_1}
    \leq
    \bra{\Phi_1}H\ket{\Phi_1}
\end{equation}
since $\ket{v_1}$ minimizes the Rayleigh quotient within the orthogonal subspace. Hence,
\begin{equation}
    \bra{\Phi_1}H\ket{\Phi_1}
    =|c_{01}|^2\lambda_0
    +|c_{11}|^2\lambda_1
    =\lambda_1
    +|c_{01}|^2
    (\lambda_0-\lambda_1) \leq \lambda_1, 
\end{equation}
as $\lambda_0<\lambda_1$ and $|c_{01}|^2 \ge 0$. As a result, one sees that  $\Lambda_1\leq\lambda_1$. This argument can be extended to any excited state by defining
\begin{equation}
    \ket{\Phi_n}
    =\sum_{m=0}^n
    c_{mn}
    \ket{\tilde{\varphi}_m}
\end{equation}
such that $\braket{v_0|\Phi_n}=\cdots=\braket{v_{n-1}|\Phi_n}=0$ and $\braket{\Phi_n|\Phi_n}=1$.  This thus completes the proof.

\subsection{Error in Estimating Excited States}

In this appendix, we analyze the error in our generalized method for estimating excited states.  
The main source of error when estimating an excited state stems from the error in the initial estimate of the ground state, so long as statistical errors are small, as in the case of the QASM simulator. 
The key issue here is whether the error in the ground state estimate may accumulate and even get magnified into the estimate of the excited states.  
For simplicity, we first consider how error propagates in applying our method to a 2-level system. We then demonstrate how the result generalizes to a 3-level system. 

Given a Hamiltonian $H$, let $\{\ket{\psi_0},\ket{\psi_1}\}$ be a complete and orthonormal set of eigenstates with $\epsilon_n\equiv\bra{\psi_n}H\ket{\psi_n}$. Suppose our variational method leads to the following approximate states 
\begin{align}
    \ket{\psi_0'}
    &=\alpha_0\ket{\psi_0}+\beta_0\ket{\psi_1}, \\
    \ket{\psi_1'}
    &=\alpha_1\ket{\psi_0}+\beta_1\ket{\psi_1}
\end{align}
as estimates of the ground and excited states with $\braket{\psi_0'|\psi_0'}=\braket{\psi_1'|\psi_1'}=1$ and $\braket{\psi_0'|\psi_1'}=0$. Without loss of generality, assume $\alpha_0$ and $\alpha_1$ are real and positive. The corresponding energy estimates are $\epsilon_n'\equiv\bra{\psi_n'}H\ket{\psi_n'}$ with errors $\delta_n\equiv\epsilon_n'-\epsilon_n$. Because the eigenstate estimates are normalized,
\begin{align}
    \delta_0
    &=|\beta_0|^2\Delta_{01}, \\
    \delta_1
    &=-\alpha_1^2\Delta_{01},
\end{align}
where $\Delta_{mn}\equiv\epsilon_n-\epsilon_m$. Additionally, orthogonality requires $-\alpha_1/\beta_0^*=\beta_1/\alpha_0$. Taking the magnitude squared gives
\begin{equation}
    \frac{\alpha_1^2}{|\beta_0|^2}
    =\frac{|\beta_1|^2}{\alpha_0^2}
    =\frac{1-\alpha_1^2}{1-|\beta_0|^2},
\end{equation}
which reveals $\alpha_1^2=|\beta_0|^2$. Thus, for a 2-level system,
\begin{equation}
    \delta_1=-\delta_0.
\end{equation}
In other words, the magnitude of the error does not change when using orthogonality to go from the ground state estimate to the excited state estimate.

For a 3-level system, let $\{\ket{\psi_0},\ket{\psi_1},\ket{\psi_2}\}$ be a complete and orthonormal set of eigenstates for $H$ and let
\begin{equation}
    \ket{\psi_n'}
    =\alpha_n\ket{\psi_0}
    +\beta_n\ket{\psi_1}
    +\gamma_n\ket{\psi_2},
    \quad
    n=1,2,3
\end{equation}
be the eigenstate estimates. Similar to the 2-level case, we assume $\alpha_0,\alpha_1,\alpha_2$ are real and positive and we derive 
\begin{align}
    \label{eq_delta0}
    \delta_0
    &=|\beta_0|^2\Delta_{01}
    +|\gamma_0|^2\Delta_{02}, \\ 
    \label{eq_delta1}
    \delta_1
    &=-\alpha_1^2\Delta_{01}
    +|\gamma_1|^2\Delta_{12}, \\
    \delta_2
    &=-\alpha_2^2\Delta_{02}
    -|\beta_2|^2\Delta_{12},
\end{align}
from normalization. Next, we minimize
\begin{equation}
    \delta_1
    =-\Delta_{01}
    +|\beta_1|^2\Delta_{01}
    +|\gamma_1|^2\Delta_{02}
\end{equation}
subject to $\alpha_0\alpha_1+\beta_0^*\beta_1+\gamma_0^*\gamma_1=0$ using the method of Lagrange multipliers. Normalization allows the orthogonality condition to be equivalently written
\begin{equation}
    0
    =g(\beta_1,\gamma_1)
    \equiv
    (1-|\gamma_0|^2)|\beta_1|^2
    +(1-|\beta_0|^2)|\gamma_1|^2
    +2\text{Re}\beta_0^*\beta_1\gamma_0\gamma_1^*
    +|\beta_0|^2+|\gamma_0|^2-1.
\end{equation}
Define the auxiliary function
\begin{equation}
    L
    \equiv
    \delta_1
    +\text{Re}(\lambda g(\beta_1,\gamma_1)),
\end{equation}
with $\lambda$ a complex lagrange multiplier. Then
\begin{align}
    0
    &=\frac{\partial L}{\partial\beta_1}
    =\beta_1^*\Delta_{01}
    +((1-|\gamma_0|^2)\beta_1^*+\beta_0^*\gamma_0\gamma_1^*)\text{Re}\lambda, \\
    0
    &=\frac{\partial L}{\partial\gamma_1}
    =\gamma_1^*\Delta_{02}
    +((1-|\beta_0|^2)\gamma_1^*+\beta_0\beta_1^*\gamma_0^*)\text{Re}\lambda.
\end{align}
Solving this set of equations gives the first excited state error:
\begin{equation}\label{eq_error1}
    \delta_1
    =-\frac{1}{2}
    \left\{
    \Delta_{01}
    -\Delta_{02}
    +\delta_0
    +\sqrt{(\Delta_{01}+\Delta_{02}-\delta_0)^2
    -4\alpha_0^2\Delta_{01}\Delta_{02}}
    \right\}.
\end{equation}
One can also minimize 
\begin{equation}
    \delta_2
    =-\Delta_{02}
    +|\beta_2|^2\Delta_{01}
    +|\gamma_2|^2\Delta_{02}
\end{equation}
subject to 
$\braket{\psi_1'|\psi_2'}=\braket{\psi_0'|\psi_2'}=0$ by introducing a second Lagrange multiplier. The method is similar, but the calculation is substantially more involved. Eventually, one finds that
\begin{equation}\label{eq_error2}
    \delta_2
    =\frac{1}{2}
    \left\{
    \Delta_{01}
    -\Delta_{02}
    -\delta_0
    +\sqrt{(\Delta_{01}+\Delta_{02}-\delta_0)^2
    -4\alpha_0^2\Delta_{01}\Delta_{02}}
    \right\}.
\end{equation}

Interestingly enough, 
\begin{equation}
    0=\delta_0+\delta_1+\delta_2,
\end{equation}
that is, the sum of the signed errors is zero, just as in the 2-level case. Additionally, one can quickly show that $\Delta_{02}>\Delta_{01}$ implies $-\delta_0\leq\delta_1\leq0$ and $-\delta_0\leq\delta_2\leq0$ using eq. (\ref{eq_delta0}) with eqs. (\ref{eq_error1}) and (\ref{eq_error2}). Evidently, the error in the ground state contributes toward an error budget that is shared between the excited states. Furthermore, $\epsilon_1'\leq\epsilon_1$ and $\epsilon_2'\leq\epsilon_2$, whereas $\epsilon_0'\geq\epsilon_0$ by the variational principle.

We now turn our attention to the values of $\delta_1$ and $\delta_2$ in a few interesting limits. First, observe that if $\alpha_0^2=1$, then $\delta_0=\delta_1=\delta_2=0$, as expected. Next, consider the almost ideal scenario where $\alpha_0^2=1-\kappa$ with $\kappa\ll1$. Using the parametrization $|\beta_0|^2=\kappa\cos^2\theta$ and $|\gamma_0|^2=\kappa\sin^2\theta$ for $0\leq\theta\leq\pi/2$,
\begin{gather}
    \delta_1
    \lesssim
    -\kappa\Delta_{01}\cos^2\theta, \\
    \delta_2\gtrsim
    -\kappa\Delta_{02}\sin^2\theta,
\end{gather}
where ``$\lesssim$" is to be read ``less than, but asymptotically equal to in the limit of small $\kappa$." To conclude this analysis, we consider the equal superposition $\alpha_0^2=|\beta_0|^2=|\gamma_0|^2=1/3$:
\begin{gather}
    \delta_0=\frac{1}{3}(\Delta_{01}+\Delta_{02}), \\
    \delta_1=-\frac{1}{3}\left(2\Delta_{01}-\Delta_{02}+\sqrt{\Delta_{01}^2+\Delta_{02}^2-\Delta_{01}\Delta_{02}}\right), \\
    \delta_2=\frac{1}{3}\left(\Delta_{01}-2\Delta_{02}+\sqrt{\Delta_{01}^2+\Delta_{02}^2-\Delta_{01}\Delta_{02}}\right).
\end{gather}
In this case, $\delta_0$, $\delta_1$, and $\delta_2$ are completely determined by the energy differences $\Delta_{01}$ and $\Delta_{02}$.

We think it is reasonable to conclude that the magnitude of the error on each excited state in a 2- or 3-level system contributed by inaccuracies in the ground state is bounded by the error on the ground state. We also think this property can be plausibly generalized for 4-level systems and beyond. Even including statistical errors on the order of $\delta_0$, the error on the excited states obtained through an orthogonalization procedure will remain comparably small. Therefore the generalized variational approach developed in this work for estimating excited states is a robust one.


















\bibliography{QCCharm_v4.bib}

\begin{thebibliography}{34}%
\makeatletter
\providecommand \@ifxundefined [1]{%
 \@ifx{#1\undefined}
}%
\providecommand \@ifnum [1]{%
 \ifnum #1\expandafter \@firstoftwo
 \else \expandafter \@secondoftwo
 \fi
}%
\providecommand \@ifx [1]{%
 \ifx #1\expandafter \@firstoftwo
 \else \expandafter \@secondoftwo
 \fi
}%
\providecommand \natexlab [1]{#1}%
\providecommand \enquote  [1]{``#1''}%
\providecommand \bibnamefont  [1]{#1}%
\providecommand \bibfnamefont [1]{#1}%
\providecommand \citenamefont [1]{#1}%
\providecommand \href@noop [0]{\@secondoftwo}%
\providecommand \href [0]{\begingroup \@sanitize@url \@href}%
\providecommand \@href[1]{\@@startlink{#1}\@@href}%
\providecommand \@@href[1]{\endgroup#1\@@endlink}%
\providecommand \@sanitize@url [0]{\catcode `\\12\catcode `\$12\catcode
  `\&12\catcode `\#12\catcode `\^12\catcode `\_12\catcode `\%12\relax}%
\providecommand \@@startlink[1]{}%
\providecommand \@@endlink[0]{}%
\providecommand \url  [0]{\begingroup\@sanitize@url \@url }%
\providecommand \@url [1]{\endgroup\@href {#1}{\urlprefix }}%
\providecommand \urlprefix  [0]{URL }%
\providecommand \Eprint [0]{\href }%
\providecommand \doibase [0]{http://dx.doi.org/}%
\providecommand \selectlanguage [0]{\@gobble}%
\providecommand \bibinfo  [0]{\@secondoftwo}%
\providecommand \bibfield  [0]{\@secondoftwo}%
\providecommand \translation [1]{[#1]}%
\providecommand \BibitemOpen [0]{}%
\providecommand \bibitemStop [0]{}%
\providecommand \bibitemNoStop [0]{.\EOS\space}%
\providecommand \EOS [0]{\spacefactor3000\relax}%
\providecommand \BibitemShut  [1]{\csname bibitem#1\endcsname}%
\let\auto@bib@innerbib\@empty
\bibitem [{\citenamefont {Cloët}\ \emph {et~al.}(2019)\citenamefont {Cloët},
  \citenamefont {Dietrich}, \citenamefont {Arrington}, \citenamefont {Bazavov},
  \citenamefont {Bishof}, \citenamefont {Freese}, \citenamefont {Gorshkov},
  \citenamefont {Grassellino}, \citenamefont {Hafidi}, \citenamefont {Jacob},
  \citenamefont {McGuigan}, \citenamefont {Meurice}, \citenamefont {Meziani},
  \citenamefont {Mueller}, \citenamefont {Muschik}, \citenamefont {Osborn},
  \citenamefont {Otten}, \citenamefont {Petreczky}, \citenamefont {Polakovic},
  \citenamefont {Poon}, \citenamefont {Pooser}, \citenamefont {Roggero},
  \citenamefont {Saffman}, \citenamefont {VanDevender}, \citenamefont {Zhang},\
  and\ \citenamefont {Zohar}}]{Cloet:2019wre}%
  \BibitemOpen
  \bibfield  {author} {\bibinfo {author} {\bibfnamefont {I.~C.}\ \bibnamefont
  {Cloët}}, \bibinfo {author} {\bibfnamefont {M.~R.}\ \bibnamefont
  {Dietrich}}, \bibinfo {author} {\bibfnamefont {J.}~\bibnamefont {Arrington}},
  \bibinfo {author} {\bibfnamefont {A.}~\bibnamefont {Bazavov}}, \bibinfo
  {author} {\bibfnamefont {M.}~\bibnamefont {Bishof}}, \bibinfo {author}
  {\bibfnamefont {A.}~\bibnamefont {Freese}}, \bibinfo {author} {\bibfnamefont
  {A.~V.}\ \bibnamefont {Gorshkov}}, \bibinfo {author} {\bibfnamefont
  {A.}~\bibnamefont {Grassellino}}, \bibinfo {author} {\bibfnamefont
  {K.}~\bibnamefont {Hafidi}}, \bibinfo {author} {\bibfnamefont
  {Z.}~\bibnamefont {Jacob}}, \bibinfo {author} {\bibfnamefont
  {M.}~\bibnamefont {McGuigan}}, \bibinfo {author} {\bibfnamefont
  {Y.}~\bibnamefont {Meurice}}, \bibinfo {author} {\bibfnamefont {Z.-E.}\
  \bibnamefont {Meziani}}, \bibinfo {author} {\bibfnamefont {P.}~\bibnamefont
  {Mueller}}, \bibinfo {author} {\bibfnamefont {C.}~\bibnamefont {Muschik}},
  \bibinfo {author} {\bibfnamefont {J.}~\bibnamefont {Osborn}}, \bibinfo
  {author} {\bibfnamefont {M.}~\bibnamefont {Otten}}, \bibinfo {author}
  {\bibfnamefont {P.}~\bibnamefont {Petreczky}}, \bibinfo {author}
  {\bibfnamefont {T.}~\bibnamefont {Polakovic}}, \bibinfo {author}
  {\bibfnamefont {A.}~\bibnamefont {Poon}}, \bibinfo {author} {\bibfnamefont
  {R.}~\bibnamefont {Pooser}}, \bibinfo {author} {\bibfnamefont
  {A.}~\bibnamefont {Roggero}}, \bibinfo {author} {\bibfnamefont
  {M.}~\bibnamefont {Saffman}}, \bibinfo {author} {\bibfnamefont
  {B.}~\bibnamefont {VanDevender}}, \bibinfo {author} {\bibfnamefont
  {J.}~\bibnamefont {Zhang}}, \ and\ \bibinfo {author} {\bibfnamefont
  {E.}~\bibnamefont {Zohar}},\ }\href@noop {} {\enquote {\bibinfo {title}
  {Opportunities for nuclear physics \& quantum information science},}\ }
  (\bibinfo {year} {2019}),\ \Eprint {http://arxiv.org/abs/1903.05453}
  {arXiv:1903.05453 [nucl-th]} \BibitemShut {NoStop}%
\bibitem [{\citenamefont {Zhang}\ \emph {et~al.}(2021)\citenamefont {Zhang},
  \citenamefont {Xing}, \citenamefont {Yan}, \citenamefont {Wang},\ and\
  \citenamefont {Zhu}}]{Zhang_2021}%
  \BibitemOpen
  \bibfield  {author} {\bibinfo {author} {\bibfnamefont {D.-B.}\ \bibnamefont
  {Zhang}}, \bibinfo {author} {\bibfnamefont {H.}~\bibnamefont {Xing}},
  \bibinfo {author} {\bibfnamefont {H.}~\bibnamefont {Yan}}, \bibinfo {author}
  {\bibfnamefont {E.}~\bibnamefont {Wang}}, \ and\ \bibinfo {author}
  {\bibfnamefont {S.-L.}\ \bibnamefont {Zhu}},\ }\href {\doibase
  10.1088/1674-1056/abd761} {\bibfield  {journal} {\bibinfo  {journal} {Chinese
  Physics B}\ }\textbf {\bibinfo {volume} {30}},\ \bibinfo {pages} {020306}
  (\bibinfo {year} {2021})}\BibitemShut {NoStop}%
\bibitem [{\citenamefont {Kharzeev}(2021)}]{Kharzeev:2021nzh}%
  \BibitemOpen
  \bibfield  {author} {\bibinfo {author} {\bibfnamefont {D.~E.}\ \bibnamefont
  {Kharzeev}},\ }\href {\doibase 10.1098/rsta.2021.0063} {\bibfield  {journal}
  {\bibinfo  {journal} {Phil. Trans. A. Math. Phys. Eng. Sci.}\ }\textbf
  {\bibinfo {volume} {380}},\ \bibinfo {pages} {20210063} (\bibinfo {year}
  {2021})},\ \Eprint {http://arxiv.org/abs/2108.08792} {arXiv:2108.08792
  [hep-ph]} \BibitemShut {NoStop}%
\bibitem [{\citenamefont {Dumitrescu}\ \emph {et~al.}(2018)\citenamefont
  {Dumitrescu}, \citenamefont {McCaskey}, \citenamefont {Hagen}, \citenamefont
  {Jansen}, \citenamefont {Morris}, \citenamefont {Papenbrock}, \citenamefont
  {Pooser}, \citenamefont {Dean},\ and\ \citenamefont
  {Lougovski}}]{Dumitrescu_2018}%
  \BibitemOpen
  \bibfield  {author} {\bibinfo {author} {\bibfnamefont {E.~F.}\ \bibnamefont
  {Dumitrescu}}, \bibinfo {author} {\bibfnamefont {A.~J.}\ \bibnamefont
  {McCaskey}}, \bibinfo {author} {\bibfnamefont {G.}~\bibnamefont {Hagen}},
  \bibinfo {author} {\bibfnamefont {G.~R.}\ \bibnamefont {Jansen}}, \bibinfo
  {author} {\bibfnamefont {T.~D.}\ \bibnamefont {Morris}}, \bibinfo {author}
  {\bibfnamefont {T.}~\bibnamefont {Papenbrock}}, \bibinfo {author}
  {\bibfnamefont {R.~C.}\ \bibnamefont {Pooser}}, \bibinfo {author}
  {\bibfnamefont {D.~J.}\ \bibnamefont {Dean}}, \ and\ \bibinfo {author}
  {\bibfnamefont {P.}~\bibnamefont {Lougovski}},\ }\href {\doibase
  10.1103/PhysRevLett.120.210501} {\bibfield  {journal} {\bibinfo  {journal}
  {Phys. Rev. Lett.}\ }\textbf {\bibinfo {volume} {120}},\ \bibinfo {pages}
  {210501} (\bibinfo {year} {2018})}\BibitemShut {NoStop}%
\bibitem [{\citenamefont {Lu}\ \emph {et~al.}(2019)\citenamefont {Lu},
  \citenamefont {Klco}, \citenamefont {Lukens}, \citenamefont {Morris},
  \citenamefont {Bansal}, \citenamefont {Ekstr\"om}, \citenamefont {Hagen},
  \citenamefont {Papenbrock}, \citenamefont {Weiner}, \citenamefont {Savage},\
  and\ \citenamefont {Lougovski}}]{Lu_2019}%
  \BibitemOpen
  \bibfield  {author} {\bibinfo {author} {\bibfnamefont {H.-H.}\ \bibnamefont
  {Lu}}, \bibinfo {author} {\bibfnamefont {N.}~\bibnamefont {Klco}}, \bibinfo
  {author} {\bibfnamefont {J.~M.}\ \bibnamefont {Lukens}}, \bibinfo {author}
  {\bibfnamefont {T.~D.}\ \bibnamefont {Morris}}, \bibinfo {author}
  {\bibfnamefont {A.}~\bibnamefont {Bansal}}, \bibinfo {author} {\bibfnamefont
  {A.}~\bibnamefont {Ekstr\"om}}, \bibinfo {author} {\bibfnamefont
  {G.}~\bibnamefont {Hagen}}, \bibinfo {author} {\bibfnamefont
  {T.}~\bibnamefont {Papenbrock}}, \bibinfo {author} {\bibfnamefont {A.~M.}\
  \bibnamefont {Weiner}}, \bibinfo {author} {\bibfnamefont {M.~J.}\
  \bibnamefont {Savage}}, \ and\ \bibinfo {author} {\bibfnamefont
  {P.}~\bibnamefont {Lougovski}},\ }\href {\doibase
  10.1103/PhysRevA.100.012320} {\bibfield  {journal} {\bibinfo  {journal}
  {Phys. Rev. A}\ }\textbf {\bibinfo {volume} {100}},\ \bibinfo {pages}
  {012320} (\bibinfo {year} {2019})}\BibitemShut {NoStop}%
\bibitem [{\citenamefont {Klco}\ \emph {et~al.}(2018)\citenamefont {Klco},
  \citenamefont {Dumitrescu}, \citenamefont {McCaskey}, \citenamefont {Morris},
  \citenamefont {Pooser}, \citenamefont {Sanz}, \citenamefont {Solano},
  \citenamefont {Lougovski},\ and\ \citenamefont {Savage}}]{Klco:2018kyo}%
  \BibitemOpen
  \bibfield  {author} {\bibinfo {author} {\bibfnamefont {N.}~\bibnamefont
  {Klco}}, \bibinfo {author} {\bibfnamefont {E.~F.}\ \bibnamefont
  {Dumitrescu}}, \bibinfo {author} {\bibfnamefont {A.~J.}\ \bibnamefont
  {McCaskey}}, \bibinfo {author} {\bibfnamefont {T.~D.}\ \bibnamefont
  {Morris}}, \bibinfo {author} {\bibfnamefont {R.~C.}\ \bibnamefont {Pooser}},
  \bibinfo {author} {\bibfnamefont {M.}~\bibnamefont {Sanz}}, \bibinfo {author}
  {\bibfnamefont {E.}~\bibnamefont {Solano}}, \bibinfo {author} {\bibfnamefont
  {P.}~\bibnamefont {Lougovski}}, \ and\ \bibinfo {author} {\bibfnamefont
  {M.~J.}\ \bibnamefont {Savage}},\ }\href {\doibase
  10.1103/PhysRevA.98.032331} {\bibfield  {journal} {\bibinfo  {journal} {Phys.
  Rev. A}\ }\textbf {\bibinfo {volume} {98}},\ \bibinfo {pages} {032331}
  (\bibinfo {year} {2018})}\BibitemShut {NoStop}%
\bibitem [{\citenamefont {Klco}\ and\ \citenamefont
  {Savage}(2019)}]{Klco:2018zqz}%
  \BibitemOpen
  \bibfield  {author} {\bibinfo {author} {\bibfnamefont {N.}~\bibnamefont
  {Klco}}\ and\ \bibinfo {author} {\bibfnamefont {M.~J.}\ \bibnamefont
  {Savage}},\ }\href {\doibase 10.1103/PhysRevA.99.052335} {\bibfield
  {journal} {\bibinfo  {journal} {Phys. Rev. A}\ }\textbf {\bibinfo {volume}
  {99}},\ \bibinfo {pages} {052335} (\bibinfo {year} {2019})}\BibitemShut
  {NoStop}%
\bibitem [{\citenamefont {Roggero}\ and\ \citenamefont
  {Carlson}(2019)}]{Roggero:2018hrn}%
  \BibitemOpen
  \bibfield  {author} {\bibinfo {author} {\bibfnamefont {A.}~\bibnamefont
  {Roggero}}\ and\ \bibinfo {author} {\bibfnamefont {J.}~\bibnamefont
  {Carlson}},\ }\href {\doibase 10.1103/PhysRevC.100.034610} {\bibfield
  {journal} {\bibinfo  {journal} {Phys. Rev. C}\ }\textbf {\bibinfo {volume}
  {100}},\ \bibinfo {pages} {034610} (\bibinfo {year} {2019})}\BibitemShut
  {NoStop}%
\bibitem [{\citenamefont {Lee}\ \emph {et~al.}(2020)\citenamefont {Lee},
  \citenamefont {Bonitati}, \citenamefont {Given}, \citenamefont {Hicks},
  \citenamefont {Li}, \citenamefont {Lu}, \citenamefont {Rai}, \citenamefont
  {Sarkar},\ and\ \citenamefont {Watkins}}]{Lee:2019zze}%
  \BibitemOpen
  \bibfield  {author} {\bibinfo {author} {\bibfnamefont {D.}~\bibnamefont
  {Lee}}, \bibinfo {author} {\bibfnamefont {J.}~\bibnamefont {Bonitati}},
  \bibinfo {author} {\bibfnamefont {G.}~\bibnamefont {Given}}, \bibinfo
  {author} {\bibfnamefont {C.}~\bibnamefont {Hicks}}, \bibinfo {author}
  {\bibfnamefont {N.}~\bibnamefont {Li}}, \bibinfo {author} {\bibfnamefont
  {B.-N.}\ \bibnamefont {Lu}}, \bibinfo {author} {\bibfnamefont
  {A.}~\bibnamefont {Rai}}, \bibinfo {author} {\bibfnamefont {A.}~\bibnamefont
  {Sarkar}}, \ and\ \bibinfo {author} {\bibfnamefont {J.}~\bibnamefont
  {Watkins}},\ }\href {\doibase 10.1016/j.physletb.2020.135536} {\bibfield
  {journal} {\bibinfo  {journal} {Phys. Lett. B}\ }\textbf {\bibinfo {volume}
  {807}},\ \bibinfo {pages} {135536} (\bibinfo {year} {2020})},\ \Eprint
  {http://arxiv.org/abs/1910.07708} {arXiv:1910.07708 [quant-ph]} \BibitemShut
  {NoStop}%
\bibitem [{\citenamefont {Lamm}\ \emph {et~al.}(2019)\citenamefont {Lamm},
  \citenamefont {Lawrence},\ and\ \citenamefont {Yamauchi}}]{Lamm:2019bik}%
  \BibitemOpen
  \bibfield  {author} {\bibinfo {author} {\bibfnamefont {H.}~\bibnamefont
  {Lamm}}, \bibinfo {author} {\bibfnamefont {S.}~\bibnamefont {Lawrence}}, \
  and\ \bibinfo {author} {\bibfnamefont {Y.}~\bibnamefont {Yamauchi}} (\bibinfo
  {collaboration} {NuQS Collaboration}),\ }\href {\doibase
  10.1103/PhysRevD.100.034518} {\bibfield  {journal} {\bibinfo  {journal}
  {Phys. Rev. D}\ }\textbf {\bibinfo {volume} {100}},\ \bibinfo {pages}
  {034518} (\bibinfo {year} {2019})}\BibitemShut {NoStop}%
\bibitem [{\citenamefont {Alexandru}\ \emph {et~al.}(2019)\citenamefont
  {Alexandru}, \citenamefont {Bedaque}, \citenamefont {Harmalkar},
  \citenamefont {Lamm}, \citenamefont {Lawrence},\ and\ \citenamefont
  {Warrington}}]{Alexandru:2019nsa}%
  \BibitemOpen
  \bibfield  {author} {\bibinfo {author} {\bibfnamefont {A.}~\bibnamefont
  {Alexandru}}, \bibinfo {author} {\bibfnamefont {P.~F.}\ \bibnamefont
  {Bedaque}}, \bibinfo {author} {\bibfnamefont {S.}~\bibnamefont {Harmalkar}},
  \bibinfo {author} {\bibfnamefont {H.}~\bibnamefont {Lamm}}, \bibinfo {author}
  {\bibfnamefont {S.}~\bibnamefont {Lawrence}}, \ and\ \bibinfo {author}
  {\bibfnamefont {N.~C.}\ \bibnamefont {Warrington}} (\bibinfo {collaboration}
  {NuQS Collaboration}),\ }\href {\doibase 10.1103/PhysRevD.100.114501}
  {\bibfield  {journal} {\bibinfo  {journal} {Phys. Rev. D}\ }\textbf {\bibinfo
  {volume} {100}},\ \bibinfo {pages} {114501} (\bibinfo {year}
  {2019})}\BibitemShut {NoStop}%
\bibitem [{\citenamefont {Ciavarella}\ \emph {et~al.}(2021)\citenamefont
  {Ciavarella}, \citenamefont {Klco},\ and\ \citenamefont
  {Savage}}]{Ciavarella:2021nmj}%
  \BibitemOpen
  \bibfield  {author} {\bibinfo {author} {\bibfnamefont {A.}~\bibnamefont
  {Ciavarella}}, \bibinfo {author} {\bibfnamefont {N.}~\bibnamefont {Klco}}, \
  and\ \bibinfo {author} {\bibfnamefont {M.~J.}\ \bibnamefont {Savage}},\
  }\href {\doibase 10.1103/PhysRevD.103.094501} {\bibfield  {journal} {\bibinfo
   {journal} {Phys. Rev. D}\ }\textbf {\bibinfo {volume} {103}},\ \bibinfo
  {pages} {094501} (\bibinfo {year} {2021})}\BibitemShut {NoStop}%
\bibitem [{\citenamefont {Atas}\ \emph {et~al.}(2021)\citenamefont {Atas},
  \citenamefont {Zhang}, \citenamefont {Lewis}, \citenamefont {Jahanpour},
  \citenamefont {Haase},\ and\ \citenamefont {Muschik}}]{Atas_2021}%
  \BibitemOpen
  \bibfield  {author} {\bibinfo {author} {\bibfnamefont {Y.~Y.}\ \bibnamefont
  {Atas}}, \bibinfo {author} {\bibfnamefont {J.}~\bibnamefont {Zhang}},
  \bibinfo {author} {\bibfnamefont {R.}~\bibnamefont {Lewis}}, \bibinfo
  {author} {\bibfnamefont {A.}~\bibnamefont {Jahanpour}}, \bibinfo {author}
  {\bibfnamefont {J.~F.}\ \bibnamefont {Haase}}, \ and\ \bibinfo {author}
  {\bibfnamefont {C.~A.}\ \bibnamefont {Muschik}},\ }\href {\doibase
  10.1038/s41467-021-26825-4} {\bibfield  {journal} {\bibinfo  {journal}
  {Nature Communications}\ }\textbf {\bibinfo {volume} {12}},\ \bibinfo {pages}
  {6499} (\bibinfo {year} {2021})}\BibitemShut {NoStop}%
\bibitem [{\citenamefont {Cohen}\ \emph {et~al.}(2021)\citenamefont {Cohen},
  \citenamefont {Lamm}, \citenamefont {Lawrence},\ and\ \citenamefont
  {Yamauchi}}]{Cohen_2021}%
  \BibitemOpen
  \bibfield  {author} {\bibinfo {author} {\bibfnamefont {T.~D.}\ \bibnamefont
  {Cohen}}, \bibinfo {author} {\bibfnamefont {H.}~\bibnamefont {Lamm}},
  \bibinfo {author} {\bibfnamefont {S.}~\bibnamefont {Lawrence}}, \ and\
  \bibinfo {author} {\bibfnamefont {Y.}~\bibnamefont {Yamauchi}} (\bibinfo
  {collaboration} {NuQS Collaboration}),\ }\href {\doibase
  10.1103/PhysRevD.104.094514} {\bibfield  {journal} {\bibinfo  {journal}
  {Phys. Rev. D}\ }\textbf {\bibinfo {volume} {104}},\ \bibinfo {pages}
  {094514} (\bibinfo {year} {2021})}\BibitemShut {NoStop}%
\bibitem [{\citenamefont {Li}\ \emph {et~al.}(2021)\citenamefont {Li},
  \citenamefont {Guo}, \citenamefont {Lai}, \citenamefont {Liu}, \citenamefont
  {Wang}, \citenamefont {Xing}, \citenamefont {Zhang},\ and\ \citenamefont
  {Zhu}}]{Li:2021kcs}%
  \BibitemOpen
  \bibfield  {author} {\bibinfo {author} {\bibfnamefont {T.}~\bibnamefont
  {Li}}, \bibinfo {author} {\bibfnamefont {X.}~\bibnamefont {Guo}}, \bibinfo
  {author} {\bibfnamefont {W.~K.}\ \bibnamefont {Lai}}, \bibinfo {author}
  {\bibfnamefont {X.}~\bibnamefont {Liu}}, \bibinfo {author} {\bibfnamefont
  {E.}~\bibnamefont {Wang}}, \bibinfo {author} {\bibfnamefont {H.}~\bibnamefont
  {Xing}}, \bibinfo {author} {\bibfnamefont {D.-B.}\ \bibnamefont {Zhang}}, \
  and\ \bibinfo {author} {\bibfnamefont {S.-L.}\ \bibnamefont {Zhu}},\
  }\href@noop {} {\enquote {\bibinfo {title} {Partonic structure by quantum
  computing},}\ } (\bibinfo {year} {2021}),\ \Eprint
  {http://arxiv.org/abs/2106.03865} {arXiv:2106.03865 [hep-ph]} \BibitemShut
  {NoStop}%
\bibitem [{\citenamefont {Kharzeev}\ and\ \citenamefont
  {Kikuchi}(2020)}]{Kharzeev:2020kgc}%
  \BibitemOpen
  \bibfield  {author} {\bibinfo {author} {\bibfnamefont {D.~E.}\ \bibnamefont
  {Kharzeev}}\ and\ \bibinfo {author} {\bibfnamefont {Y.}~\bibnamefont
  {Kikuchi}},\ }\href {\doibase 10.1103/PhysRevResearch.2.023342} {\bibfield
  {journal} {\bibinfo  {journal} {Phys. Rev. Res.}\ }\textbf {\bibinfo {volume}
  {2}},\ \bibinfo {pages} {023342} (\bibinfo {year} {2020})},\ \Eprint
  {http://arxiv.org/abs/2001.00698} {arXiv:2001.00698 [hep-ph]} \BibitemShut
  {NoStop}%
\bibitem [{\citenamefont {Tu}\ \emph {et~al.}(2020)\citenamefont {Tu},
  \citenamefont {Kharzeev},\ and\ \citenamefont {Ullrich}}]{Tu:2019ouv}%
  \BibitemOpen
  \bibfield  {author} {\bibinfo {author} {\bibfnamefont {Z.}~\bibnamefont
  {Tu}}, \bibinfo {author} {\bibfnamefont {D.~E.}\ \bibnamefont {Kharzeev}}, \
  and\ \bibinfo {author} {\bibfnamefont {T.}~\bibnamefont {Ullrich}},\ }\href
  {\doibase 10.1103/PhysRevLett.124.062001} {\bibfield  {journal} {\bibinfo
  {journal} {Phys. Rev. Lett.}\ }\textbf {\bibinfo {volume} {124}},\ \bibinfo
  {pages} {062001} (\bibinfo {year} {2020})},\ \Eprint
  {http://arxiv.org/abs/1904.11974} {arXiv:1904.11974 [hep-ph]} \BibitemShut
  {NoStop}%
\bibitem [{\citenamefont {Kitaev}(1995)}]{kitaev1995quantum}%
  \BibitemOpen
  \bibfield  {author} {\bibinfo {author} {\bibfnamefont {A.~Y.}\ \bibnamefont
  {Kitaev}},\ }\href@noop {} {\enquote {\bibinfo {title} {Quantum measurements
  and the abelian stabilizer problem},}\ } (\bibinfo {year} {1995}),\ \Eprint
  {http://arxiv.org/abs/quant-ph/9511026} {arXiv:quant-ph/9511026 [quant-ph]}
  \BibitemShut {NoStop}%
\bibitem [{\citenamefont {Abrams}\ and\ \citenamefont
  {Lloyd}(1999)}]{PhysRevLett.83.5162}%
  \BibitemOpen
  \bibfield  {author} {\bibinfo {author} {\bibfnamefont {D.~S.}\ \bibnamefont
  {Abrams}}\ and\ \bibinfo {author} {\bibfnamefont {S.}~\bibnamefont {Lloyd}},\
  }\href {\doibase 10.1103/PhysRevLett.83.5162} {\bibfield  {journal} {\bibinfo
   {journal} {Phys. Rev. Lett.}\ }\textbf {\bibinfo {volume} {83}},\ \bibinfo
  {pages} {5162} (\bibinfo {year} {1999})}\BibitemShut {NoStop}%
\bibitem [{\citenamefont {O'Brien}\ \emph {et~al.}(2019)\citenamefont
  {O'Brien}, \citenamefont {Tarasinski},\ and\ \citenamefont
  {Terhal}}]{OBrien_2019}%
  \BibitemOpen
  \bibfield  {author} {\bibinfo {author} {\bibfnamefont {T.~E.}\ \bibnamefont
  {O'Brien}}, \bibinfo {author} {\bibfnamefont {B.}~\bibnamefont {Tarasinski}},
  \ and\ \bibinfo {author} {\bibfnamefont {B.~M.}\ \bibnamefont {Terhal}},\
  }\href {\doibase 10.1088/1367-2630/aafb8e} {\bibfield  {journal} {\bibinfo
  {journal} {New J. Phys.}\ }\textbf {\bibinfo {volume} {21}},\ \bibinfo
  {pages} {023022} (\bibinfo {year} {2019})}\BibitemShut {NoStop}%
\bibitem [{\citenamefont {Peruzzo}\ \emph {et~al.}(2014)\citenamefont
  {Peruzzo}, \citenamefont {McClean}, \citenamefont {Shadbolt}, \citenamefont
  {Yung}, \citenamefont {Zhou}, \citenamefont {Love}, \citenamefont
  {Aspuru-Guzik},\ and\ \citenamefont {O’Brien}}]{Peruzzo_2014}%
  \BibitemOpen
  \bibfield  {author} {\bibinfo {author} {\bibfnamefont {A.}~\bibnamefont
  {Peruzzo}}, \bibinfo {author} {\bibfnamefont {J.}~\bibnamefont {McClean}},
  \bibinfo {author} {\bibfnamefont {P.}~\bibnamefont {Shadbolt}}, \bibinfo
  {author} {\bibfnamefont {M.-H.}\ \bibnamefont {Yung}}, \bibinfo {author}
  {\bibfnamefont {X.-Q.}\ \bibnamefont {Zhou}}, \bibinfo {author}
  {\bibfnamefont {P.~J.}\ \bibnamefont {Love}}, \bibinfo {author}
  {\bibfnamefont {A.}~\bibnamefont {Aspuru-Guzik}}, \ and\ \bibinfo {author}
  {\bibfnamefont {J.~L.}\ \bibnamefont {O’Brien}},\ }\href {\doibase
  10.1038/ncomms5213} {\bibfield  {journal} {\bibinfo  {journal} {Nature
  Communications}\ }\textbf {\bibinfo {volume} {5}},\ \bibinfo {pages} {4213}
  (\bibinfo {year} {2014})}\BibitemShut {NoStop}%
\bibitem [{\citenamefont {McClean}\ \emph {et~al.}(2016)\citenamefont
  {McClean}, \citenamefont {Romero}, \citenamefont {Babbush},\ and\
  \citenamefont {Aspuru-Guzik}}]{McClean_2016}%
  \BibitemOpen
  \bibfield  {author} {\bibinfo {author} {\bibfnamefont {J.~R.}\ \bibnamefont
  {McClean}}, \bibinfo {author} {\bibfnamefont {J.}~\bibnamefont {Romero}},
  \bibinfo {author} {\bibfnamefont {R.}~\bibnamefont {Babbush}}, \ and\
  \bibinfo {author} {\bibfnamefont {A.}~\bibnamefont {Aspuru-Guzik}},\ }\href
  {\doibase 10.1088/1367-2630/18/2/023023} {\bibfield  {journal} {\bibinfo
  {journal} {New Journal of Physics}\ }\textbf {\bibinfo {volume} {18}},\
  \bibinfo {pages} {023023} (\bibinfo {year} {2016})}\BibitemShut {NoStop}%
\bibitem [{\citenamefont {Shen}\ \emph {et~al.}(2017)\citenamefont {Shen},
  \citenamefont {Zhang}, \citenamefont {Zhang}, \citenamefont {Zhang},
  \citenamefont {Yung},\ and\ \citenamefont {Kim}}]{Shen_2017}%
  \BibitemOpen
  \bibfield  {author} {\bibinfo {author} {\bibfnamefont {Y.}~\bibnamefont
  {Shen}}, \bibinfo {author} {\bibfnamefont {X.}~\bibnamefont {Zhang}},
  \bibinfo {author} {\bibfnamefont {S.}~\bibnamefont {Zhang}}, \bibinfo
  {author} {\bibfnamefont {J.-N.}\ \bibnamefont {Zhang}}, \bibinfo {author}
  {\bibfnamefont {M.-H.}\ \bibnamefont {Yung}}, \ and\ \bibinfo {author}
  {\bibfnamefont {K.}~\bibnamefont {Kim}},\ }\href {\doibase
  10.1103/PhysRevA.95.020501} {\bibfield  {journal} {\bibinfo  {journal} {Phys.
  Rev. A}\ }\textbf {\bibinfo {volume} {95}},\ \bibinfo {pages} {020501}
  (\bibinfo {year} {2017})}\BibitemShut {NoStop}%
\bibitem [{\citenamefont {Bali}(2001)}]{Bali:2000gf}%
  \BibitemOpen
  \bibfield  {author} {\bibinfo {author} {\bibfnamefont {G.~S.}\ \bibnamefont
  {Bali}},\ }\href {\doibase 10.1016/S0370-1573(00)00079-X} {\bibfield
  {journal} {\bibinfo  {journal} {Physics Reports}\ }\textbf {\bibinfo {volume}
  {343}},\ \bibinfo {pages} {1} (\bibinfo {year} {2001})}\BibitemShut {NoStop}%
\bibitem [{\citenamefont {Kawanai}\ and\ \citenamefont
  {Sasaki}(2012)}]{Kawanai:2011jt}%
  \BibitemOpen
  \bibfield  {author} {\bibinfo {author} {\bibfnamefont {T.}~\bibnamefont
  {Kawanai}}\ and\ \bibinfo {author} {\bibfnamefont {S.}~\bibnamefont
  {Sasaki}},\ }\href {\doibase 10.1103/PhysRevD.85.091503} {\bibfield
  {journal} {\bibinfo  {journal} {Phys. Rev. D}\ }\textbf {\bibinfo {volume}
  {85}},\ \bibinfo {pages} {091503} (\bibinfo {year} {2012})}\BibitemShut
  {NoStop}%
\bibitem [{\citenamefont {Tilly}\ \emph {et~al.}(2021)\citenamefont {Tilly},
  \citenamefont {Chen}, \citenamefont {Cao}, \citenamefont {Picozzi},
  \citenamefont {Setia}, \citenamefont {Li}, \citenamefont {Grant},
  \citenamefont {Wossnig}, \citenamefont {Rungger}, \citenamefont {Booth},\
  and\ \citenamefont {Tennyson}}]{Tilly_2021}%
  \BibitemOpen
  \bibfield  {author} {\bibinfo {author} {\bibfnamefont {J.}~\bibnamefont
  {Tilly}}, \bibinfo {author} {\bibfnamefont {H.}~\bibnamefont {Chen}},
  \bibinfo {author} {\bibfnamefont {S.}~\bibnamefont {Cao}}, \bibinfo {author}
  {\bibfnamefont {D.}~\bibnamefont {Picozzi}}, \bibinfo {author} {\bibfnamefont
  {K.}~\bibnamefont {Setia}}, \bibinfo {author} {\bibfnamefont
  {Y.}~\bibnamefont {Li}}, \bibinfo {author} {\bibfnamefont {E.}~\bibnamefont
  {Grant}}, \bibinfo {author} {\bibfnamefont {L.}~\bibnamefont {Wossnig}},
  \bibinfo {author} {\bibfnamefont {I.}~\bibnamefont {Rungger}}, \bibinfo
  {author} {\bibfnamefont {G.~H.}\ \bibnamefont {Booth}}, \ and\ \bibinfo
  {author} {\bibfnamefont {J.}~\bibnamefont {Tennyson}},\ }\href {\doibase
  10.48550/ARXIV.2111.05176} {\enquote {\bibinfo {title} {The variational
  quantum eigensolver: a review of methods and best practices},}\ } (\bibinfo
  {year} {2021})\BibitemShut {NoStop}%
\bibitem [{\citenamefont {Hylleraas}\ and\ \citenamefont
  {Undheim}(1930)}]{HU1930}%
  \BibitemOpen
  \bibfield  {author} {\bibinfo {author} {\bibfnamefont {E.~A.}\ \bibnamefont
  {Hylleraas}}\ and\ \bibinfo {author} {\bibfnamefont {B.}~\bibnamefont
  {Undheim}},\ }\href {\doibase 10.1007/BF01397263} {\bibfield  {journal}
  {\bibinfo  {journal} {Zeitschrift f{\"u}r Physik}\ }\textbf {\bibinfo
  {volume} {65}},\ \bibinfo {pages} {759} (\bibinfo {year} {1930})}\BibitemShut
  {NoStop}%
\bibitem [{\citenamefont {MacDonald}(1933)}]{PhysRev.43.830}%
  \BibitemOpen
  \bibfield  {author} {\bibinfo {author} {\bibfnamefont {J.~K.~L.}\
  \bibnamefont {MacDonald}},\ }\href {\doibase 10.1103/PhysRev.43.830}
  {\bibfield  {journal} {\bibinfo  {journal} {Phys. Rev.}\ }\textbf {\bibinfo
  {volume} {43}},\ \bibinfo {pages} {830} (\bibinfo {year} {1933})}\BibitemShut
  {NoStop}%
\bibitem [{\citenamefont {Anand}\ \emph {et~al.}(2022)\citenamefont {Anand},
  \citenamefont {Schleich}, \citenamefont {Alperin-Lea}, \citenamefont
  {Jensen}, \citenamefont {Sim}, \citenamefont {D{\'{\i}}az-Tinoco},
  \citenamefont {Kottmann}, \citenamefont {Degroote}, \citenamefont
  {Izmaylov},\ and\ \citenamefont {Aspuru-Guzik}}]{Anand_2022}%
  \BibitemOpen
  \bibfield  {author} {\bibinfo {author} {\bibfnamefont {A.}~\bibnamefont
  {Anand}}, \bibinfo {author} {\bibfnamefont {P.}~\bibnamefont {Schleich}},
  \bibinfo {author} {\bibfnamefont {S.}~\bibnamefont {Alperin-Lea}}, \bibinfo
  {author} {\bibfnamefont {P.~W.~K.}\ \bibnamefont {Jensen}}, \bibinfo {author}
  {\bibfnamefont {S.}~\bibnamefont {Sim}}, \bibinfo {author} {\bibfnamefont
  {M.}~\bibnamefont {D{\'{\i}}az-Tinoco}}, \bibinfo {author} {\bibfnamefont
  {J.~S.}\ \bibnamefont {Kottmann}}, \bibinfo {author} {\bibfnamefont
  {M.}~\bibnamefont {Degroote}}, \bibinfo {author} {\bibfnamefont {A.~F.}\
  \bibnamefont {Izmaylov}}, \ and\ \bibinfo {author} {\bibfnamefont
  {A.}~\bibnamefont {Aspuru-Guzik}},\ }\href {\doibase 10.1039/d1cs00932j}
  {\bibfield  {journal} {\bibinfo  {journal} {Chemical Society Reviews}\
  }\textbf {\bibinfo {volume} {51}},\ \bibinfo {pages} {1659} (\bibinfo {year}
  {2022})}\BibitemShut {NoStop}%
\bibitem [{\citenamefont {Jordan}\ and\ \citenamefont
  {Wigner}(1928)}]{JW_1928}%
  \BibitemOpen
  \bibfield  {author} {\bibinfo {author} {\bibfnamefont {P.}~\bibnamefont
  {Jordan}}\ and\ \bibinfo {author} {\bibfnamefont {E.}~\bibnamefont
  {Wigner}},\ }\href {\doibase 10.1007/BF01331938} {\bibfield  {journal}
  {\bibinfo  {journal} {Z. Physik}\ }\textbf {\bibinfo {volume} {47}},\
  \bibinfo {pages} {631} (\bibinfo {year} {1928})}\BibitemShut {NoStop}%
\bibitem [{\citenamefont {Giurgica-Tiron}\ \emph {et~al.}(2020)\citenamefont
  {Giurgica-Tiron}, \citenamefont {Hindy}, \citenamefont {LaRose},
  \citenamefont {Mari},\ and\ \citenamefont {Zeng}}]{Giurgica_Tiron_2020}%
  \BibitemOpen
  \bibfield  {author} {\bibinfo {author} {\bibfnamefont {T.}~\bibnamefont
  {Giurgica-Tiron}}, \bibinfo {author} {\bibfnamefont {Y.}~\bibnamefont
  {Hindy}}, \bibinfo {author} {\bibfnamefont {R.}~\bibnamefont {LaRose}},
  \bibinfo {author} {\bibfnamefont {A.}~\bibnamefont {Mari}}, \ and\ \bibinfo
  {author} {\bibfnamefont {W.~J.}\ \bibnamefont {Zeng}},\ }in\ \href {\doibase
  10.1109/QCE49297.2020.00045} {\emph {\bibinfo {booktitle} {2020 IEEE
  International Conference on Quantum Computing and Engineering (QCE)}}}\
  (\bibinfo {year} {2020})\ pp.\ \bibinfo {pages} {306--316}\BibitemShut
  {NoStop}%
\bibitem [{\citenamefont {Wang}\ and\ \citenamefont
  {Zunger}(1994)}]{Wang_1994}%
  \BibitemOpen
  \bibfield  {author} {\bibinfo {author} {\bibfnamefont {L.}~\bibnamefont
  {Wang}}\ and\ \bibinfo {author} {\bibfnamefont {A.}~\bibnamefont {Zunger}},\
  }\href {\doibase 10.1063/1.466486} {\bibfield  {journal} {\bibinfo  {journal}
  {J. Chem. Phys.}\ }\textbf {\bibinfo {volume} {100}},\ \bibinfo {pages}
  {2394} (\bibinfo {year} {1994})},\ \Eprint
  {http://arxiv.org/abs/https://doi.org/10.1063/1.466486}
  {https://doi.org/10.1063/1.466486} \BibitemShut {NoStop}%
\bibitem [{\citenamefont {Santagati}\ \emph {et~al.}(2018)\citenamefont
  {Santagati}, \citenamefont {Wang}, \citenamefont {Gentile}, \citenamefont
  {Paesani}, \citenamefont {Wiebe}, \citenamefont {McClean}, \citenamefont
  {Morley-Short}, \citenamefont {Shadbolt}, \citenamefont {Bonneau},
  \citenamefont {Silverstone}, \citenamefont {Tew}, \citenamefont {Zhou},
  \citenamefont {O’Brien},\ and\ \citenamefont {Thompson}}]{Santagati_2018}%
  \BibitemOpen
  \bibfield  {author} {\bibinfo {author} {\bibfnamefont {R.}~\bibnamefont
  {Santagati}}, \bibinfo {author} {\bibfnamefont {J.}~\bibnamefont {Wang}},
  \bibinfo {author} {\bibfnamefont {A.~A.}\ \bibnamefont {Gentile}}, \bibinfo
  {author} {\bibfnamefont {S.}~\bibnamefont {Paesani}}, \bibinfo {author}
  {\bibfnamefont {N.}~\bibnamefont {Wiebe}}, \bibinfo {author} {\bibfnamefont
  {J.~R.}\ \bibnamefont {McClean}}, \bibinfo {author} {\bibfnamefont
  {S.}~\bibnamefont {Morley-Short}}, \bibinfo {author} {\bibfnamefont {P.~J.}\
  \bibnamefont {Shadbolt}}, \bibinfo {author} {\bibfnamefont {D.}~\bibnamefont
  {Bonneau}}, \bibinfo {author} {\bibfnamefont {J.~W.}\ \bibnamefont
  {Silverstone}}, \bibinfo {author} {\bibfnamefont {D.~P.}\ \bibnamefont
  {Tew}}, \bibinfo {author} {\bibfnamefont {X.}~\bibnamefont {Zhou}}, \bibinfo
  {author} {\bibfnamefont {J.~L.}\ \bibnamefont {O’Brien}}, \ and\ \bibinfo
  {author} {\bibfnamefont {M.~G.}\ \bibnamefont {Thompson}},\ }\href {\doibase
  10.1126/sciadv.aap9646} {\bibfield  {journal} {\bibinfo  {journal} {Science
  Advances}\ }\textbf {\bibinfo {volume} {4}},\ \bibinfo {pages} {eaap9646}
  (\bibinfo {year} {2018})}\BibitemShut {NoStop}%
\bibitem [{\citenamefont {Lee}\ \emph {et~al.}(2019)\citenamefont {Lee},
  \citenamefont {Huggins}, \citenamefont {Head-Gordon},\ and\ \citenamefont
  {Whaley}}]{Lee_2019}%
  \BibitemOpen
  \bibfield  {author} {\bibinfo {author} {\bibfnamefont {J.}~\bibnamefont
  {Lee}}, \bibinfo {author} {\bibfnamefont {W.~J.}\ \bibnamefont {Huggins}},
  \bibinfo {author} {\bibfnamefont {M.}~\bibnamefont {Head-Gordon}}, \ and\
  \bibinfo {author} {\bibfnamefont {K.~B.}\ \bibnamefont {Whaley}},\ }\href
  {https://doi.org/10.1021/acs.jctc.8b01004} {\bibfield  {journal} {\bibinfo
  {journal} {J. Chem. Theory Comput.}\ }\textbf {\bibinfo {volume} {15}},\
  \bibinfo {pages} {311} (\bibinfo {year} {2019})}\BibitemShut {NoStop}%
\end{thebibliography}%

\end{document}